\documentclass[12pt]{article}
\usepackage{latexsym,graphicx,a4,epsfig,psfrag,here,color}

\begin{document}
\renewcommand{\theequation}{\arabic{section}.\arabic{equation}}
\newcommand{\mathbold}[1]{\mbox{\boldmath $\bf#1$}}
\newcommand{\eq}{\begin{eqnarray}}
\newcommand{\en}{\end{eqnarray}}
\newcommand{\NChPT}{ChPT$\,^{\!\bigcirc\hspace{-0.6em}\raisebox{0.021em}
{\tiny  A}\hspace{0.2em}}$~}

\thispagestyle{empty}

\title{{\normalsize {\tt \hspace*{6.cm} Preprints: ECT*-03-28, HISKP-TH-03/22}}\\[6mm]
Chiral Perturbation Theory in a Nuclear Background}

\author{
L. Girlanda~$^{\rm a}$, A.~Rusetsky~$^{\rm a,b,c}$, and 
W. Weise~$^{\rm a,d}$ \\[2mm]
{\em{\small $^{\rm a}$ECT*, Villa Tambosi, Strada delle Tabarelle 286, 
I-38050}}\\[-2mm]
{\em{\small Villazzano (Trento), Italy}}\\
{\em{\small $^{\rm b}$Universit\"{a}t Bonn, Helmholtz-Institut f\"{u}r
Strahlen- und Kernphysik (Theorie),}}
\\[-2mm] 
{\em{\small Nu\ss allee 14-16, D-53115 Bonn, Germany}}\\
{\em{\small $^{\rm c}$On leave of absence from High Energy Physics Institute,
Tbilisi State University}} \\[-2mm]
{\em{\small University St.~9, 380086 Tbilisi, Georgia}}\\
{\em{\small $^{\rm d}$Physik-Department, Technische Universit\"at M\"{u}nchen,}} 
\\[-2mm]
{\em{\small D-85747 Garching, Germany}}\\
}

\maketitle

\begin{abstract}

We propose a novel way to formulate Chiral Perturbation Theory in a nuclear 
background, characterized by a static, non-uniform distribution of the
baryon number that describes the {\em finite} nucleus. In the limiting case of a
uniform distribution, the theory reduces to the well-known
zero-temperature in-medium ChPT. The proposed approach is used to calculate
the self-energy of the charged pion in the background of the heavy
nucleus at $O(p^5)$ in the chiral expansion, and to derive the leading
terms of the  pion-nucleus optical potential.

\vspace*{.1cm}

\noindent
PACS number(s): 11.10.St, 12.39.Fe, 21.65.+f, 36.10.Gv

\end{abstract}

\thispagestyle{empty}


\newpage

\setcounter{equation}{0}
\section{Introduction}
\label{sec:intro}

The topic of low-energy pion-nucleus interactions has a long history and a
well-developed phenomenology~\cite{Ericson,EW,Thomas}. Energy spectra of pionic
atoms have traditionally played an important role in promoting the 
understanding of this field by providing a large and systematic data base
which sets tight constraints on the parameterization of the pion-nuclear potential 
(see Ref.~\cite{potential} for a state-of-the-art update).
 
New developments, both experimentally and theoretically, have recently 
revitalized this field. Accurate measurements of deeply bound (1$s$) states 
of pionic atoms formed with $Pb$ and $Sn$ isotopes~\cite{experiments} have 
triggered renewed interest in the underlying mechanisms governing $S$-wave 
pion-nucleus interactions. 
The quest for  ``fingerprints of chiral restoration'' in this context 
has been raised~\cite{Weise:sg,Yamazaki}, referring to a possible in-medium 
reduction of the pion decay constant which is an order parameter of the
spontaneously broken chiral symmetry in QCD~\cite{fpi}.
Recent theoretical investigations~\cite{KW,Oller,Annals,Korea} based on Chiral 
Perturbation Theory (ChPT) have, for one part, focused on the calculation 
of the in-medium shift of the pion mass, to be identified with the value of the threshold 
$S$-wave $\pi$-nucleus optical potential in the center of the nucleus. 
Alternatively, the energy-dependent pion self-energy (polarization function) 
has been extrapolated, using ChPT input and
the local density approximation, to calculate directly the pionic $1s$ and
$2p$ level shifts and widths for $Pb$ and $Sn$ isotopes~\cite{Kolo}.

What all these recent investigations have in common, is their emphasis on a 
framework based on chiral dynamics and on the spontaneously broken chiral 
symmetry of low-energy QCD.
However, all these analyses rely, in one way or another, on assumptions 
which need further systematic scrutiny and clarification:

\begin{itemize}
\item[i)]
The in-medium mass of the pion is a well-defined quantity within the 
in-medium ChPT. The procedure of extracting the empirical value for this 
quantity, however, involves detailed theoretical analysis based on 
phenomenological potentials, introducing model dependence which is hard to 
control.

\item[ii)]
The wave 
function of a deeply bound pionic state is concentrated near the surface of 
the nucleus,
where the variation of the proton and fermion density distributions is 
maximal~\cite{Soff}. 
The commonly used expansion in gradients of the local nuclear density must 
therefore be carefully examined.

\item[iii)]
Pion-nuclear observables should not depend on the parameterization 
of the interpolating pion field in the chiral Lagrangian.
Yet, questions concerning off-shell ambiguity have appeared repeatedly in 
recent works. In Ref.~\cite{Korea}
it is demonstrated that the off-shell ambiguity disappears in the mass
(as it should of course), if one considers the systematic chiral expansion of 
the in-medium
mass shift. It remains to prove the equivalent statement
for the bound-state 
mass spectrum, e.g. as obtained in the approach of Ref.~\cite{Kolo}.
Last but not least, 
the complete expression of the
pion self-energy operator at $O(p^5)$ and $O(p^6)$, including
all strong and electromagnetic isospin-breaking effects at this orders, cannot be found in the literature 
so far.

\end{itemize}

It is therefore evident that further progress
in the description of pion-nuclear 
bound systems requires the formulation of the systematic
ChPT framework in the presence of a {\em finite} nucleus.
This is what our present paper is focused on.
The interest in this problem is motivated by at least three reasons:

\begin{itemize}

\item[i)]
It has been argued (e.g.~\cite{Kolo}) that chiral dynamics
(the specific energy dependence of low-energy pion-nucleus interactions as 
imposed by chiral symmetry, in combination with the approximate 
vanishing of the isospin-even $\pi N$ amplitude) is
a key ingredient in understanding the empirically observed ``missing 
repulsion'' in the $S$-wave pion-nucleus optical potential. This statement 
has to be verified on a more systematic basis.

\item[ii)]
The in-medium mass shift of the charged pion is the only quantity known to us,
which is related to the two-point Green function and 
where the $O(p^2)$ $\pi N$ electromagnetic low-energy constant (LEC) 
$f_1$ enters at lowest order
independently from other unknown LECs. On the other hand, this 
constant is very important in the 
theoretical analysis~\cite{EPJC} of the experimental results obtained by
 the Pionic 
Hydrogen Collaboration (PSI)~\cite{PSI} (in particular, it provides the
bulk of uncertainty in the relation of the ground-state energy-level shift
to the $S$-wave $\pi N$ scattering lengths~\cite{EPJC}). One can expect
that reliable 
experimental information on the in-medium mass shift can, at least,
test the order-of-magnitude estimates for $f_1$, on which the current 
numerical analysis of pionic hydrogen is based~\cite{EPJC,Meissnerf1,Faessler}.

\item[iii)]
The consistent formulation of ChPT in the background of a finite nucleus is a 
challenging task of its own. It opens perspectives  for many applications of 
ChPT in nuclear physics -- especially for the problems
in which the finite boundaries of the system are crucial.

\end{itemize}

The problem that we are addressing can be formulated as follows.
At low energy the interactions between pions, photons and nucleons are
described 
by the conventional Lagrangian of ChPT 
(we concentrate here exclusively on the 2-flavor case). 
Let $\Omega$ be the bound state of a nucleus 
containing $A$ nucleons. We aim to describe
the processes $\Omega+X\to \Omega'+Y$ where $X,Y$ represent any number
of pions and photons (in this paper we consider the case where
both $X$ and $Y$ stand for  $1$-pion states). In principle one can generalize the approach to cases 
such as pion inelastic scattering or pion absorption, where unbound nucleons
are also present.
This is, however, beyond the scope of the present paper.

A rigorous field theoretical treatment would require solving the
nuclear bound-state (e.g. Bethe-Salpeter) equation and then
calculating transition matrix elements with in- and out-going pion fields. This
task is simplified considerably for heavy nuclei (in the large-$A$
limit). In particular, in this limit the recoil of the nucleus as a whole is a $1/A$
correction and can be neglected, and one may formally set $\Omega'=\Omega$. 
ChPT in a nuclear background which will 
be referred to as ``Chiral Perturbation Theory for Heavy Nuclei'' (\NChPT, 
where ${\!\bigcirc\hspace{-0.83em}\raisebox{0.01em}
{\small  A}\hspace{0.2em}}$
denotes the heavy nucleus with the mass number $A$) hereafter, 
is then to be understood as an approximate theory in which the transitions
$\Omega+X\to \Omega+Y$ are treated in terms of simpler Green functions
``in the presence of the nucleus'', $G_\Omega(X\to Y)$. The process
$X\to Y$ is  described by chiral dynamics, and the presence of the
nucleus $\Omega$ is  parametrized in terms of a few phenomenologically 
determined functions (e.g. proton and neutron distributions)
which are directly related to corresponding observables.

In-medium ChPT for a homogeneous system starts from a nuclear Fermi
gas and systematically introduces interactions mediated by real or
virtual pions. \NChPT  is a generalization for finite
systems. It reduces to the standard in-medium ChPT in the limit of a uniform
distribution of baryon number. Beyond that, \NChPT provides a systematic
framework to study pion-nuclear bound states, for which the finite
volume and the surface of the nucleus are important ingredients.

\NChPT imposes chiral counting rules not only on the hierarchy
of pion-nucleon interactions, but also on the relevant nuclear matrix
elements. The nuclear structure information required is thus limited
to expectation values of only those operators whose chiral dimension
is compatible with the given order in the chiral counting at which
the calculations are carried out. 
We demonstrate explicitly how this works in the calculation of the
pion self-energy at  $O(p^5)$ in ChPT. To that chiral order, the
complete set of leading terms in pion-nucleus optical potential (those
linear in the proton and neutron densities, $\rho_p({\bf r})$ and
$\rho_n({\bf r})$) are generated. It is of course well known from
pion-nuclear phenomenology \cite{Ericson, EW, Thomas}, that rescattering
and absorption terms of $O(p^6)$ and higher orders are quantitatively
important. In the present paper we focus on the systematic expansion
to $O(p^5)$. Further developments concerning higher orders are planned
for subsequent publications.

The layout of the paper is as follows. In section \ref{sec:framework}
we consider in detail the construction of the framework: main assumptions,
role of chiral symmetry and the connection to the conventional in-medium
ChPT. In section \ref{sec:Pi}, using the formulated framework, we
obtain the complete 
expression for the pion self-energy operator at $O(p^5)$ in ChPT in the 
background of the finite nucleus. Using the latter expressions  the
pion-nucleus optical potential at $O(p^5)$ is derived in section
\ref{sec:bound}.   
In section \ref{sec:comp} we compare with the existing approaches in
the literature. 
Conclusions are drawn in section \ref{sec:cncl}.

\setcounter{equation}{0}
\section{The framework}
\label{sec:framework}
\subsection{$S$-matrix as functional of the free fields}
In order to ensure that the Green functions
 satisfy
constraints imposed by chiral symmetry at every step of the
calculation, and do not depend on the  parameterization of the pion field,
we equip the effective Lagrangian with external $c$-number sources
and consider the generating function in the presence of these sources.
Thus, our approach is closely related to the approach of 
Refs.~\cite{Oller,Annals} which uses exactly the same setting. 
The difference arises at the following points:
\begin{itemize}
\item
The main difference is that our approach is not limited to the uniform baryon
density distribution and takes the finite-size effects into account. In the 
limit of the uniform distribution, our approach will reduce to that of
Refs.~\cite{Oller,Annals}.
\item
The electromagnetic effects are not considered in~\cite{Oller,Annals}, although the
existing framework allows for accommodating them without further modification.
Since in the description of the pionic atoms where our approach will be 
applied, the Coulomb interaction plays an important role, we include the 
electromagnetic effects from the beginning. 
\item
Neglecting 4- and more-fermion interactions in Refs.~\cite{Oller,Annals}
leads to substantial technical simplifications. In our approach, we do not
make this approximation in the perturbative expansion of the generating functional. Note that  in ChPT there is no 
contribution from the 4-fermion interactions to the pion nucleus optical 
potential at $O(p^5)$ (see below). For the treatment of the multinucleon
interactions in the kinematical region where $NN$ pairs are close to threshold, in addition, 
some kind of non-perturbative resummation might be necessary due to the large $NN$ scattering length \cite{Annals}.
\end{itemize}
The formulation which was considered in Refs.~\cite{KW,Korea,Kolo} does not 
use external sources. For this reason, in the uniform-density limit
it is equivalent to our approach only on the mass shell, i.e. in the 
calculation of the physical observables. Moreover, since 
Refs.~\cite{KW,Korea,Kolo} are based on the standard HBChPT description of
pion-nucleon interactions~\cite{HBChPT}, the above equivalence holds up to
the contributions from the non-standard counting regime (see below)\footnote{
Throughout the paper we use the terminology of Ref. \cite{Annals} for 
different momentum regimes in the Feynman integrals}.
Finally note, that our approach which is based on constructing the 
$S$-operator in the perturbation theory and then sandwiching it by the 
state vectors describing particles in the medium, is formally similar 
to the one used in Ref. \cite{Mallik} -- again, in the case of the uniform medium.

We start with setting up the shorthand notation for the external sources
in the generating functional
\eq\label{j} 
j\doteq\{ s,p,v,a\}\, ,\quad\quad
j_0\doteq\{ {\cal M}_q,0,0,0\}\, ,
\en
where $s,p,v,a$ stand for scalar, pseudoscalar, vector and axial-vector
external sources, and ${\cal M}_q={\rm diag}\, (m_u,m_d)$ denotes the quark mass
matrix. The Green functions are obtained from the generating functional
by differentiating with respect to the pertinent external sources, and setting
$j=j_0$ at the end.

The Lagrangian which we shall use to describe the system, is the standard 
Lagrangian of ChPT with pions, nucleons, photons and external sources 
in the vacuum:
\eq\label{lag}
{\cal L}={\cal L}_\gamma+{\cal L}_\pi^{(p^2)}+{\cal L}_\pi^{(e^2)}
+{\cal L}_\pi^{(p^4)}+{\cal L}_\pi^{(e^2p^2)}+\cdots+
{\cal L}_N^{(p)}+{\cal L}_N^{(p^2)}+{\cal L}_N^{(e^2)}+\cdots\, ,
\en
where the dots stand for the higher-order terms in the chiral expansion, 
for the terms of order $e^4$ and higher that are consistently neglected from 
now on, and for the terms in the
Lagrangian containing four and more nucleon fields. For the convenience of the
reader we collect the known expressions for the lowest-order chiral 
Lagrangians in appendix \ref{app:lagr}.

The asymptotic ($in$- and $out$-) states of the theory include both
``elementary'' 
particles (nucleons/antinucleons, pions, photons) as well as bound 
systems of nucleons (stable nuclei). On the other hand, 
 bound systems of pions with nuclei --  pionic atoms --
are not present in the set of the asymptotic states, because in the present 
study their width can not be neglected. Instead, the pionic atoms will show up
as  (complex) poles on the second Riemann sheet in the pertinent Green 
functions. 
In this section, however,
we consider the states containing only ``elementary'' particles.
 For example, the state $|{\bf n;\bar n;m;k};in(out)\rangle$ with
$n$ nucleons, $\bar n$ antinucleons, $m$ pions and $k$ photons, is
given by
\eq\label{state}
|{\bf p}_1s_1,\cdots {\bf p}_ns_n;\bar{\bf p}_1\bar s_1,\cdots 
\bar{\bf p}_{\bar n}\bar s_{\bar n};
{\bf q}_1,\cdots{\bf q}_m;
{\bf l}_1\varepsilon_1,\cdots {\bf l}_k\varepsilon_k;in(out)\rangle\, .
\en
Here, $s_i$ ($\bar s_i$) stands for the spin 
projection of the nucleon (antinucleon), 
and $\varepsilon_i$ for the photon polarization 
vector. To ease notations, we suppress the indices that distinguish 
the different species of the nucleons (antinucleons) and pions. 

The generic off-shell transition process
$X\to Y$ in the presence of any number of elementary fermions and bosons 
in $in$- and $out$-states, with the sets $X,Y$ both containing
only bosons (pions and photons), is described in ChPT by the following
matrix element
\eq\label{ME}
\langle {\bf n';\bar n';m';k'};out|
T{\bf o}_1(y_1)\cdots {\bf o}_\omega(y_\omega) {\bf o}_1(x_1)\cdots {\bf o}_\rho(x_\rho)
|{\bf n;\bar n;m;k};in\rangle\, .
\en
Here two sets of bilinear quark currents
with pertinent quantum numbers (pseudoscalar current for pions, 
vector current for photons): $ {\bf o}_1(x_1)\cdots {\bf o}_\rho(x_\rho)$ and 
${\bf o}_1(y_1)\cdots {\bf o}_\omega(y_\omega)$ describe the bosons present in $X$ and
$Y$, respectively.

The expression (\ref{ME}) is given in the Heisenberg picture. We further
wish to rewrite
this expression in the interaction picture. This can be done 
in a standard manner, using the
evolution operator (see, e.g. \cite{Schweber}). The result is given by
\eq\label{ME_intrc}
\langle n';\bar n';m';k'|
To_1(y_1)\cdots o_\omega(y_\omega) o_1(x_1)\cdots o_\rho(x_\rho)\, {\bf S}
|n;\bar n;m;k\rangle\, ,
\en
where ${\bf S}$ denotes the conventional $S$-operator, 
$o_1(y_1)\cdots o_\omega(y_\omega) o_1(x_1)\cdots o_\rho(x_\rho)$ stand for the currents
in the interaction picture, and the free-particle states $|n;\bar n;m;k\rangle$
are the eigenstates of the unperturbed Hamiltonian. Using 
Wick's theorem, the $T$-product in Eq.~(\ref{ME_intrc}) can be expressed in terms
of the normal products of free nucleon, pion and photon fields
\eq\label{freefields}
\Psi(x)&=&\pmatrix{\Psi_p(x)\cr\Psi_n(x)}\, ,\quad\quad
\Phi(x)\,=\,\pmatrix{\Phi_+(x)\cr\Phi_0(x)\cr\Phi_-(x)}\, ,
\nonumber\\[2mm]
\Psi_i(x)&=&\sum_s\int\frac{d^3{\bf k}}{(2\pi)^32\sqrt{M_i^2+{\bf k}}}\,
({\rm e}^{-ik\cdot x}b_i({\bf k},s)u_i({\bf k},s)+
{\rm e}^{ik\cdot x}d_i^\dagger({\bf k},s)v_i({\bf k},s))\, ,
\nonumber\\[2mm]
\Phi_\alpha(x)&=&\int\frac{d^3{\bf k}}{(2\pi)^32\sqrt{m_\alpha^2+{\bf k}^2}}\,
(a_\alpha({\bf k}){\rm e}^{-ik\cdot x}
+a_\alpha^\dagger({\bf k}){\rm e}^{ik\cdot x})\, ,
\nonumber\\[2mm]
{\cal A}_\mu(x)&=&\sum_\lambda\int\frac{d^3{\bf k}}{(2\pi)^32|{\bf k}|}\,
(c_\lambda({\bf k})\varepsilon_\mu(\lambda){\rm e}^{-ik\cdot x}
+c_\lambda^\dagger({\bf k})\varepsilon_\mu^*(\lambda){\rm e}^{ik\cdot x})\, .
\en
Here, $i=p,n$ and $\alpha=\pm,0$ stand for different nucleon and pion species.
In the subsequent expressions we shall suppress these indices, if this does 
not lead to the confusion.
Furthermore, $u_i({\bf k},s)$, $v_i({\bf k},s)$ denote conventional free
Dirac spinors
and $\varepsilon_\mu(\lambda)$ is the photon polarization vector.
The state vectors from Eq. (\ref{ME_intrc}) are obtained by repeated action
of creation operators on the perturbative vacuum
\eq\label{nnbmk}
|n;\bar n;m;k\rangle=\frac{1}{\sqrt{n!\bar n!m!k!}}
\underbrace{b^\dagger({\bf p},s)\cdots}_{n}
\underbrace{d^\dagger(\bar{\bf p},\bar s)\cdots}_{\bar n}
\underbrace{a^\dagger({\bf k})\cdots}_{m}
\underbrace{c^\dagger({\bf l})\cdots}_{k}|0\rangle\, .
\en

In the canonical formalism, the free fields introduced above coincide
with the interpolating field operators at time $t=0$. At an arbitrary $t$,
the relation between free and interpolating fields is given by the
evolution operator. However, in order to avoid imposing  
boundary conditions on the interpolating pion field which is not
uniquely  defined, we reformulate the framework
 in terms of the functional integral depending on
the external sources $j$. To this end, we consider the generating functional
in the theory given by Lagrangian~(\ref{lag})
\eq\label{Z}
Z(j|\eta,\bar\eta)=\int {\cal D}U{\cal D}\psi{\cal D}\bar\psi
{\cal D}A_\mu\,
\exp\biggl\{i\int({\cal L}+\bar\eta\psi+\bar\psi\eta)d^4\!x\biggr\}\, ,
\en
where
${\cal L}={\cal L}(U,\psi,\bar\psi,A_\mu;j)$. Here $U$ stands for the 
pion interpolating field matrix, $A_\mu$ is the photon field, 
$\psi=\pmatrix{\psi_p\cr\psi_n}$ 
denotes the two-component interpolating nucleon field and $\bar\psi=\psi^\dagger \gamma_0$ its  conjugate. $\eta=\pmatrix{\eta_p\cr\eta_n}$ is the nucleon 
external source and $\bar\eta$ its conjugate. 
Note that
the meson Green functions which are obtained by differentiating the quantity
$Z$ with respect to the bosonic external sources $j$, do not depend on the parameterization of the
pion interpolating field $U$, the latter being merely an integration variable in
the functional integral. Thus, the ``off-shell ambiguity'' mentioned
in the introduction, never arises in this formulation.

At the first step, we expand the generating functional (\ref{Z}), 
in powers of nucleon sources $\eta,\bar\eta$
\eq\label{Zeta}
\hspace*{-.5cm}&&Z(j|\eta,\bar\eta)=Z^{(0)}(j)+
\int d^4\!x d^4\!y\, \bar\eta(x) Z^{(1)}(j|x,y)\eta(y)
\nonumber\\[2mm]
\hspace*{-.5cm}&+&\frac{1}{2}\,\int d^4\!x_1 d^4\!x_2 d^4\!y_1 d^4\!y_2\, 
\bar\eta(x_1)\bar\eta(x_2) Z^{(2)}(j|x_1,x_2,y_1,y_2)\eta(y_1)\eta(y_2)
+\cdots\, .
\en
(Note that,
since the interactions with the external sources $j$ preserve the baryon 
number, we always have equal number of $\eta$'s and $\bar\eta$'s in this
expansion.)

We further define the operator $\hat S(j)$ through the following construction:
\eq\label{SFock}
\hspace*{-1.1cm}&&
\hat S(j)=S^{(0)}(j)
+
\int d^4\!x d^4\!y :\bar\Psi(x) S^{(1)}(j|x,y)\Psi(y):
\nonumber\\[2mm]
\hspace*{-1.1cm}&+&\!\!\!\!\!
\frac{1}{2}\int d^4\!x_1 d^4\!x_2 d^4\!y_1 d^4\!y_2 :\bar\Psi(x_1)\bar\Psi(x_2) 
S^{(2)}(j|x_1,x_2,y_1,y_2)\Psi(y_1)\Psi(y_2):
+\cdots\, ,
\en
where
\eq\label{SZ}
S^{(0)}(j)&=&Z^{(0)}(j)\, ,
\nonumber\\[2mm]
S^{(1)}(j|x,y)&=&z_N^{-1/2}\stackrel{\longrightarrow}{(i\not\!\partial_x-M)}
Z^{(1)}(j|x,y)\stackrel{\longleftarrow}{(-i\not\!\partial_y-M)}z_N^{-1/2}\, ,
\en
and so forth.
Here $M={\rm diag}\, (M_p,M_n)$ and $z_N={\rm diag}\,(z_p,z_n)$ are,
respectively, the 
$2\times 2$ nucleon mass matrix and the nucleon wave function renormalization 
matrix. 
Furthermore, ``$:(\cdots ):$'' denotes the normal ordering of the
operators. 
The operator $\hat S(j)$ has the meaning of
the $S$-matrix defined on the subspace of the state vectors containing only fermions, in the presence of the external bosonic sources $j$.

In order to obtain the matrix element (\ref{ME_intrc}), we further
differentiate the operator $\hat S(j)$ with respect to the pertinent
bosonic sources, and expand the result in powers of $j$.
This result is symbolically written as
\eq\label{Sj} 
&&\frac{(-i)^{\rho+\omega}\delta^{\rho+\omega}\hat S(j)}
{\delta j_1(y_1)\cdots\delta j_\omega(y_\omega)\delta j_1(x_1)\cdots\delta j_\rho(x_\rho)}
=\hat S_0^{\{Y,X\}}(y_1\cdots y_\omega;x_1\cdots x_\rho)
\nonumber\\[3mm]
&\!\!+\!\!&\int d^4z j_\sigma(z)\hat S_1^{\{Y,X\},\sigma}(y_1\cdots y_\omega;x_1\cdots x_\rho;z)
\nonumber\\[2mm]
&\!\!+\!\!&\frac{1}{2}\,\int d^4z_1 d^4z_2 j_\sigma(z_1)j_\delta(z_2)
\hat S_2^{\{Y,X\},\sigma\delta}(y_1\cdots y_\omega;x_1\cdots x_\rho;z_1,z_2)+\cdots\, ,
\en
where the index of the external source $j$ corresponds to the quantum numbers
of the quark current: $j_\sigma\doteq s,p,v,a$.
Further, in a complete analogy with Eqs. (\ref{Zeta}), (\ref{SFock}) and
(\ref{SZ}), we construct the operator
\eq\label{barS}
\hspace*{-.2cm}\bar S^{\{Y,X\}}(y_1\cdots y_\omega;x_1\cdots x_\rho)&\!\!=\!\!&
\hat S_0^{\{Y,X\}}(y_1\cdots y_\omega;x_1\cdots x_\rho)
\nonumber\\[2mm]
\hspace*{-3.5cm}&&\hspace*{-1.6cm}+\,\,\int d^4z :\Phi_\alpha(z): 
\bar S_1^{\{Y,X\},\alpha}(y_1\cdots y_\omega;x_1\cdots x_\rho;z)\hspace*{1.cm}
\nonumber\\[2mm]
\hspace*{-3.5cm}&&\hspace*{-1.6cm}+\,\,\int d^4z :{\cal A}_\mu(z):
 \bar S_1^{\{Y,X\},\mu}(y_1\cdots y_\omega;x_1\cdots x_\rho;z)
+\cdots\, ,\hspace*{1.cm}
\en
where, in analogy with Eq. (\ref{SZ}), the coefficient
$\bar S^{\{Y,X\}\cdots}$ in (\ref{barS}) are obtained from 
$\hat S^{\{Y,X\}\cdots}$ appearing in Eq. (\ref{Sj}):
e.g., $\bar S^{\{Y,X\}\alpha}(y_1\cdots y_\omega;x_1\cdots x_\rho;z)$ is obtained
from $\hat S^{\{Y,X\}\alpha}(y_1\cdots y_\omega;x_1\cdots x_\rho;z)$ by removing
the pion pole in the Fourier transform with respect to the variable $z$, and
 multiplying the result by the wave-function renormalization factor 
$\sqrt{z_{\pi^\alpha}}$. 
Integrating with the solution of the free-field equation
$\Phi_\alpha(z)$ finally puts the 4-momentum, conjugate to the
variable $z$, on   
the mass shell. Other coefficients in the expansion (\ref{barS}) are obtained 
similarly. We do not repeat their (evident) explicit expressions
here. At the end, the operator $\bar S^{\{Y,X\}}(y_1\cdots
y_\omega;x_1\cdots x_\rho)$  
is expressed in terms of the normal products of all free fields 
$:\bar\Psi\cdots \Phi_\alpha\cdots{\cal A}_\mu\cdots\Psi:$ 
(we recall that $\hat S^{\{Y,X\}\cdots}(y_1\cdots y_\omega;x_1\cdots x_\rho;\cdots)$
are given by infinite series over normal products of fermion fields,
cf. with Eq. (\ref{SFock})). 

Finally, the matrix element of the $T$-product of the currents in the 
Heisenberg picture (\ref{ME}) can be rewritten in the form of a matrix 
element of the operator $\bar S^{\{Y,X\}}$ between the free-particle states
\eq\label{ME_int}
\langle n';\bar n';m';k'| 
\bar S^{\{Y,X\}}(y_1\cdots y_\omega;x_1\cdots x_\rho)|n;\bar n;m,k\rangle\, .
\en
This is the result we were looking for.
The following remarks are in order:
\begin{itemize}

\item[i)]
The construction presented in this section is  the
conventional LSZ formalism for the calculation of the matrix elements 
(\ref{ME}) in the language of functional integrals (see e.g. \cite{Faddeev}). 
Note that here 
one does not impose boundary conditions on the interpolating fields at $t=0$.
This is crucial for ensuring chiral symmetry at every step, and for 
circumventing the problem of the ``off-shell ambiguity''.

\item[ii)]
The coefficient functions of the operators $\hat S(j)$ and
$\bar S^{\{Y,X\}}$  are ultraviolet-finite, since the effective Lagrangian
(\ref{lag}) includes the set of counterterms that make all $S$-matrix
elements finite. On the other hand, $Z(j|\eta,\bar\eta)$, in general,
can not be made ultraviolet-finite with chirally-invariant counterterms 
alone \cite{BL2}.

\item[iii)]
In the presence of photons, the coefficient functions in
$\hat S(j)$ and $\bar S^{\{Y,X\}}$ are infrared-divergent.
We shall regularize the infrared (as well as the ultraviolet) divergences
by using the dimensional regularization. The mass-shell limit is then understood
in space-time dimensions different from four.

\item[iv)]
The coefficient functions in $\hat S(j)$ and $\bar S^{\{Y,X\}}$ are, 
in general, gauge-de\-pend\-ent, since they involve off-shell legs. Only
those quantities that correspond to physically  
observable processes should be gauge-independent.

\item[v)]
In the calculation of the coefficient functions, some prescription is 
implied that preserves chiral power counting in the presence of nucleons.
For example, one could have chosen heavy baryon
ChPT~\cite{HBChPT}, or the infrared regularization method~\cite{BL}. 
As the most general prescription, we propose to use 
the so-called threshold expansion
of the relativistic Feynman integrals~\cite{Beneke}
which, by construction, allows to keep the contributions from all
soft integration regimes. In particular, the 
threshold expansion will automatically
produce the contributions to the Feynman integrals not only from the 
``HBChPT regime'', but from the ``non-standard regime(s)'' \cite{Annals} as
well. Below we assume that the threshold 
expansion is always applied to all Feynman diagrams, even if this is not 
explicitly stated.

\end{itemize}

\subsection{Scattering processes in the presence of the nucleus}

The asymptotic spectrum of the theory described by the Lagrangian~(\ref{lag})
includes one-particle state vectors representing the nucleus as a bound
system of $A$ nucleons: 
\eq\label{bound}
|\mathbold{\Omega};in\rangle=|\mathbold{\Omega};out\rangle
=|\mathbold{\Omega}\rangle\, ,\quad\quad
\langle\mathbold{\Omega}'|\mathbold{\Omega}\rangle=
(2\pi)^3\, 2P_\Omega^0\,\,\delta^3({\bf P}'_\Omega-{\bf P}_\Omega)\, .
\en

At this stage $|{\bf \Omega}\rangle$ can still be any highly excited
nuclear state as well.
The scattering process $\Omega+X\to\Omega'+Y$ is described by the following 
matrix element:
\eq\label{ME_Omega}
&&\langle \mathbold{\Omega}'|
T{\bf o}_1(y_1)\cdots {\bf o}_\omega(y_\omega) {\bf o}_1(x_1)\cdots {\bf o}_\rho(x_\rho)
|\mathbold{\Omega}\rangle
\nonumber\\[2mm]
&=&\langle \Omega'|
To_1(y_1)\cdots o_\omega(y_\omega) o_1(x_1)\cdots o_\rho(x_\rho) {\bf S}
|\Omega\rangle\, .
\en
Here
\eq\label{Omega_intrc}
|\Omega\rangle={\bf U}(-\infty,0)|\mathbold{\Omega}\rangle\, ,\quad\quad
\langle\Omega'|=\langle\mathbold{\Omega}'|{\bf U}(0,+\infty)\, ,
\en
where ${\bf U}(t',t)$ stands for the conventional evolution operator
(see, e.g. \cite{Schweber}).

The basis $|n;\bar n;m;k\rangle$ of the eigenstates of the free Hamiltonian
is complete.
A general nuclear state $|\Omega\rangle$ can be written as a linear combination of
the basis states
\eq\label{linear}
|\Omega\rangle=\sum_{n;\bar n;m;k}C_{n;\bar n;m;k}|{n;\bar n;m;k}\rangle\, .
\en
From the comparison to the expression of the bound-state matrix elements
in the Mandelstam formalism, one can obtain the relation to the 
Bethe-Salpeter amplitude for the bound state
\eq\label{equaltime}
C_{n;\bar n;m;k}=\int d^3{\bf x}_1\cdots d^3{\bf x}_A 
G_{n;\bar n;m;k}({\bf x}_1\cdots {\bf x}_A)
\langle 0|\mathbold{\Psi}(0,{\bf x}_1)\cdots\mathbold{\Psi}(0,{\bf x}_A)|
\mathbold{\Omega}\rangle\, ,
\en
where $\mathbold{\Psi}(t,{\bf x})$ denotes the nucleon field operator in the 
Heisenberg representation. To make the interpretation easier,
we found it useful to give this relation in a form containing 
the equal-time amplitude, which in the non-relativistic limit reduces to the 
Schr\"{o}dinger wave function of the nucleus.
In the language that uses the external sources,
the Bethe-Salpeter amplitude is defined via the residues of the pertinent Green 
functions, without referring to the matrix elements of the interpolating
fields. Further, $G_{n;\bar n;m;k}({\bf x}_1\cdots {\bf x}_A)$
stands for the perturbative kernel, which can, in principle, be calculated in 
ChPT. However, in our approach one does not need the explicit knowledge of
neither this kernel, nor the Bethe-Salpeter amplitude.

The quantity (\ref{ME_Omega}) which we want to calculate is given in the form
of a matrix element of the operator $\bar S^{\{YX\}}$ 
introduced in the previous section, between the nuclear state vectors
(\ref{linear})
\eq\label{ME_Sbar}
\hspace*{-.5cm}&&\hspace*{-.5cm}\langle \mathbold{\Omega}'|
T{\bf o}_1(y_1)\cdots {\bf o}_\omega(y_\omega) {\bf o}_1(x_1)\cdots {\bf o}_\rho(x_\rho)
|\mathbold{\Omega}\rangle
=\langle \Omega'|
\bar S^{\{YX\}}(y_1\cdots y_\omega,x_1\cdots x_\rho)|\Omega\rangle\, .
\nonumber\\
&&
\en
The right-hand side of this equation has the form
$\sum_i S_i\langle \Omega'|:P_i:|\Omega\rangle$, where $S_i$ denote the
coefficient functions, and $:P_i:$ stand for the normal products of the free
fields $:\bar\Psi\cdots \Phi_\alpha\cdots{\cal A}_\mu\cdots\Psi:$.
The main idea of \NChPT consists in the following: one calculates the 
coefficient functions $S_i$ in standard ChPT, whereas the matrix elements 
of the normal products of the free fields are parametrized in terms of a few 
characteristic functions describing nuclear dynamics. 
This is the place where the empirical input enters in our 
calculational scheme\footnote{
As it stands, the approach is based on a well-defined separation of the 
perturbative and non-perturbative contributions in the matrix elements of
currents. Using the same arguments one could, in principle, reshuffle 
particular contributions between these two groups, and design a different 
scheme. Here, we shall not explore this possibility any further.}.
Our strategy consists in relating these matrix 
elements to a limited number of nuclear observables, and predicting
other observables where the same matrix elements occur.

If the approach is used to describe a large nucleus ($A\to\infty$), in
its ground state, the following assumption can be justified:

{\em Assumption 1.} Matrix elements of operators containing at least
one free pion or photon field, vanish between the nuclear states.
We interpret this property in the following way:
 $|\Omega\rangle$ describes the ground state of the nucleus as a
system of nucleons only at $A\to \infty$.

The above assumption enables one to rewrite
the matrix element (\ref{ME_Omega}) in terms of the operator $\hat S(j)$
defined by Eq. (\ref{SFock}), as
\eq\label{ME_Shar}
\langle \Omega'|\frac{(-i)^{\rho+\omega}\delta^{\rho+\omega}\hat S(j)}
{\delta j_1(y_1)\cdots\delta j_\omega(y_\omega)\delta j_1(x_1)\cdots\delta j_\rho(x_\rho)}|\Omega\rangle\biggr|_{j=j_0}\, .
\en

{\em Assumption 2.}
In order to simplify the present  calculations, we assume that the
nucleus is  in the spin-saturated state. This means
that the nucleus is not characterized by internal quantum numbers.
The latter assumption, which can obviously be
relaxed on a later stage, in combination with the static limit ($A\to \infty$), helps to reduce significantly -- on the basis of
symmetry considerations alone -- the amount of empirical input needed on
the nuclear side.

{\em Assumption 3.}
Finally, we assume that the nuclear
background does not affect the ultraviolet properties of the theory.
In the case of the infinite nuclear medium, this is an exact statement,
since all diagrams with medium insertions contain a cutoff at Fermi-momentum
$k_F$. The generalization to finite systems amounts to require that
Fourier transforms of distributions involving $|\Omega\rangle$ drop
sufficiently rapidly at large momentum.
Put differently, we assume that in the expansion 
$\sum_i S_i\langle \Omega'|:P_i:|\Omega\rangle$ each matrix element is finite,
and there are no cancellations of divergences between various terms
(we recall that $S_i$ are finite after renormalization).

It should be emphasized that, with the above assumptions, \NChPT
can be pursued to any given chiral order. The consistency of the
approach, and the validity of the above
assumptions, should then be verified {\em a posteriori}
order by order in the chiral expansion.

\subsection{Pion-nucleus bound state}
\label{sec:boundstate}

The primary aim of the present paper is to study states of negatively charged 
pions bound to heavy nuclei. To this end, it is useful to consider the two-point function of the pseudoscalar  densities, sandwiched by the
nuclear states
\eq\label{def-D}
\bar D_{\Omega'\Omega}(q^\prime,q)=\frac{-i}{(2BF)^2}\,\int d^4\!x
d^4\!y\,{\rm e}^{iq^\prime \cdot   
x-iq\cdot y}\,
\langle\Omega'|\frac{\delta^2 \hat S(j)}
{\delta p^+(x)\delta p^-(y)}|\Omega\rangle\biggr|_{j=j_0}\, ,
\en
where $F$ is the pion decay constant in the chiral limit, $B$ is
related to the quark condensate ($F^2 B = -\lim_{m_u, m_d \to 0}
\langle \bar q q \rangle$).
We follow the Condon-Shortley phase convention and express
the isovector (pseudoscalar) source field as 
$p^\pm(x)=(2)^{-1/2}(\mp p_1(x)+ip_2(x)),~p^0(x)=p_3(x)$. The normalization
in Eq.~(\ref{def-D}) is defined so that in lowest order of the
chiral expansion,
the residue in the pole of this two-point function is normalized to unity.

Using translational invariance,  we remove 
an overall $\delta$-function, corresponding
to the conservation of the total 3-momentum, from the quantity
$\bar D_{\Omega'\Omega}(q^\prime,q)$. In the remaining matrix
element one may pass to the static limit 
$P_\Omega'=P_\Omega=M_\Omega(1,{\bf 0})$.
In the calculations, it is convenient to use the 
Hilbert space where the static nucleus is described by the 
state vector $|\Omega)$ normalized to unity \footnote{
This state vector, which describes the nucleus in the static limit,
is in fact defined by Eq. (\ref{def-D1}). The more detailed treatment
of the static limit is considered below, in section 
\ref{sec:recoil}.}. Using  the new notations,
the right-hand side of Eq. (\ref{def-D}) at $A\to\infty$ can be written as
\eq\label{def-D1}
&&\bar D_{\Omega'\Omega}(q^\prime,q)=(2\pi)^3\delta^3({\bf P}_\Omega'+{\bf q^\prime}-
{\bf P}_\Omega-{\bf q})\bar D(q^\prime,q)\, ,
\nonumber\\[2mm]
&&\bar D(q^\prime,q)=
\frac{-i}{(2BF)^2}\,\int d^4\!x d^4\!y\,{\rm e}^{iq^\prime \cdot
x-iq\cdot y}\,
(\Omega|\frac{\delta^2 \hat S(j)}
{\delta p^+(x)\delta p^-(y)}|\Omega)\biggr|_{j=j_0}\, ,
\en
For a static nuclear background, energy conservation implies
\eq\label{DbarD}
\bar D(q^\prime,q)=2\pi\delta({q^0}^\prime-q^0)D(E;{\bf q^\prime},{\bf q})\, ,\quad\quad
E={q^0}^\prime=q^0\, .
\en
The pion self-energy is defined from the inverse of
the operator $D$ as follows:
\eq\label{Pi}
\Pi(E;{\bf q^\prime},{\bf q})=(2\pi)^3
\delta^3({\bf q^\prime}-{\bf q})D_0^{-1}(E;{\bf q}^2)-D^{-1}(E;{\bf q^\prime},{\bf q})\, ,
\en
where $D_0^{-1}(E;{\bf q}^2)=E^2 -{\bf q}^2 - m_{\pi}^2$ is the inverse
of the free pion propagator, and $m_{\pi}\doteq m_{\pi^+}$ 
stands for the physical mass of the charged pion in the vacuum.

Our strategy for finding the energy levels and widths of the pionic 
bound state implies the following steps:
\begin{enumerate}
\item
Calculate the non-local 
pion self-energy operator 
$\Pi(E;{\bf q^\prime},{\bf q})$ in the systematic ChPT  expansion.
\item
Find the pole position in the two-point function $D(E;{\bf q^\prime},{\bf q})$
with a given self-energy operator $\Pi(E;{\bf q^\prime},{\bf q})$, using 
non-perturbative methods. The real and imaginary parts 
of the pole position are related to the real energy and decay 
width of this state. 
\item
At the last step, for a given self-energy operator calculated perturbatively 
at a fixed order in ChPT, 
study the chiral expansion of the energy and the decay width.
This enables one to control the convergence in the bound-state characteristics
at all stages of calculations,
possibly eliminating (potentially large) higher-order contributions that arise
due to the use of the non-perturbative techniques.
\end{enumerate}

It should be understood that identifying the (complex) bound-state energy with
the pole position of the pion two-point function in the nuclear background
(\ref{DbarD}) implies additional assumptions:

{\em Assumption 4.}
In Eq. (\ref{def-D}) the nucleus is on shell in its ground state, whereas 
in the pion-nucleus bound state both pion and nucleus are off shell.
The rigorous definition of the bound-state energy involves finding the
pole position of the Green function of $2A$ fermion and two pion fields.
We expect, however, that for a very large $A$ one may safely put the 
nucleus on mass shell.

{\em Assumption 5.} 
If the nucleus is put on shell, this means that the effect coming from 
nuclear excited 
states is neglected. In order to demonstrate this, note that the 
initial-/final- state interactions in the $A$ ingoing/outgoing nucleon 
legs of any Green 
function create a tower of radial-excited nuclear bound states. 
Putting the nuclear momentum on the mass shell eliminates all but the 
ground state contribution to the Green function and thus, from the 
mathematical viewpoint, amounts to "freezing" the nucleus in its ground 
state during the whole interaction with the projectile 
\cite{Foldy,Agassi}.

By construction, the energy of the bound state obtained with this procedure,
does not have any spurious dependence on the choice of parameterization 
for the pion field $U$. If, instead of the quantity $\bar D$ given by 
Eq.~(\ref{def-D}), one considers the two-point function of the interpolating 
pion field, such a dependence arises in this 
two-point function itself~\cite{Korea,Kolo}. As one would expect, this
dependence indeed disappears in the pole position up to a given order in 
the chiral expansion~\cite{Korea}. When solving the bound-state equation 
numerically for the non-local self-energy~\cite{Kolo}, it is in general 
difficult to 
demonstrate that the dependence on the parameterization of the pion field
is of higher chiral order. It is therefore  advantageous 
to work in a framework which avoids this dependence {\em ab initio}
[of course, the pole positions in the two-point functions of the pseudoscalar 
densities, and of the pion fields on the other hand, 
should coincide order by order in the chiral expansion.].

We finally note that the two-point function is gauge-dependent.
In order to demonstrate that the approach is consistent at a given chiral 
order, one has to explicitly show that the pole position is invariant up
to this order.
In this paper, we check the gauge invariance up to and including
 $O(p^5)$ (see section~\ref{sec:Pi}). 
At present, 
we are however not aware of the general proof of the gauge invariance that
extends to all orders within our approach.
If it turns out at higher orders, 
that the gauge invariance of the bound-state energy is lost, this would signal
about the inadequacy of the approximations which were used in the formulation
of the present framework.
From the phenomenological point of view, the issue is totally irrelevant,
since it concerns only the higher-order 
electromagnetic corrections to the pion-nuclear optical
potential.

\subsection{Free Fermi-gas}

In order to examine the limit of a homogeneous nuclear medium, it
 is instructive to start with the simple case of a Fermi gas. The 
ground state of this system has all levels filled with nucleons up to the 
Fermi-momentum $k_F$.
To simplify notations, we assume here in addition 
that isospin symmetry is not broken, switch off the electromagnetic 
interactions, and suppress the index labeling different
nucleon species $p$ and $n$.
The ground state in this approximation is
\eq\label{Fermi}
|\Omega\rangle\rightarrow|\Omega_0\rangle=
{\cal N}\,\prod^{k_F}_{{\bf l}_i,s_i=\pm 1/2} b^\dagger({\bf l}_i,s_i)|0\rangle\, ,
\en
where $|0\rangle$ denotes the perturbative vacuum, and  ${\cal N}$ is a
normalization constant. Furthermore, we again introduce the ``round-bracket'' 
notation, with $( \Omega_0 | \Omega_0 )=1$ [see section \ref{sec:recoil} for 
further details.].

Using the canonical commutation relations, one can easily show that
\eq\label{commutators}
&&(\Omega_0|
b^\dagger({\bf l}_1,s_1)b({\bf l}_2,s_2)|\Omega_0)=
(2\pi)^3\, 2l_1^0\, \delta^3({\bf l}_1-{\bf l}_2)\delta_{s_1s_2}\,\theta(k_F-|{\bf l}_1|)\, ,
\nonumber\\[2mm]
&&(\Omega_0|
b^\dagger({\bf l}_1,s_1)b^\dagger({\bf l}_2,s_2)b({\bf l}_3,s_3)
b({\bf l}_4,s_4)|\Omega_0)
\nonumber\\[2mm]
&=&(2\pi)^3\, 2l_1^0\, \delta^3({\bf l}_1-{\bf l}_4)\delta_{s_1s_4}\,\theta(k_F-|{\bf l}_1|)\,\, (2\pi)^3\, 2l_2^0\, 
\delta^3({\bf l}_2-{\bf l}_3)\delta_{s_2s_3}\,\theta(k_F-|{\bf  l}_2|)
\nonumber\\[2mm]
&-&[perm.~1\leftrightarrow 2]\, ,
\en
and so forth. The free fermion (nucleon) propagator in momentum space, with
$\Omega_0$ in the background, is given by
\eq\label{1prop}
&&S_{\Omega_0}(p)=
i\int d^4\!x {\rm e}^{ip\cdot x}\,(\Omega_0|T\Psi(x)\bar\Psi(0)
|\Omega_0)
\nonumber\\[2mm]
&=&
i\int d^4\!x {\rm e}^{ip\cdot x}\,\langle 0|T\Psi(x)\bar\Psi(0)|0\rangle
+i\int d^4\!x {\rm e}^{ip\cdot x}\,
(\Omega_0|:\Psi(x)\bar\Psi(0):|\Omega_0)
\nonumber\\[2mm]
&=&(M_\Psi+\not\! p)\biggl(\frac{1}{M_\Psi^2-p^2-i0}
-2\pi i\delta(p^2-M_\Psi^2)\theta(p^0)\theta(k_F-|{\bf p}|)\biggr)\, ,
\en
where $M_\Psi$  stands for the mass of the fermion field. One immediately 
recognizes that the quantity $S_{\Omega_0}(p)$ is the relativistic 
in-medium free fermion propagator.

\begin{figure}[t]
\begin{center}
\includegraphics[width=5.cm]{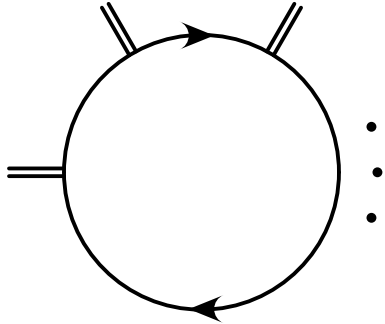}
\end{center}
\caption{Closed fermion loop with any number of vertices $\Gamma_1,~\Gamma_2,\cdots$. 
Double lines denote the external sources $j$ attached to the loop.}
\label{fig:closedloop}
\end{figure}

In the calculation of the Green functions of the external sources
$j$, one needs to evaluate only the closed fermion lines at
non-zero fermion density (see Fig.~\ref{fig:closedloop}). 
Let us first consider the closed loop with two vertices $\Gamma_{1,2}$,
where these stand for arbitrary matrices in the space of Dirac and
flavor indices, describing the coupling of the external sources
$j$ to the fermion bilinears.
Applying Wick's theorem and using Eq.~(\ref{commutators}) 
for the normal products, we get   
\eq\label{Wick}
&&i^2
(\Omega_0|T\bar\Psi(x)\Gamma_1\Psi(x)\,\bar\Psi(y)\Gamma_2\Psi(y)
|\Omega_0)
\nonumber\\[2mm]
&=&-{\rm Tr}\, \bigl[S_{\Omega_0}(x-y)\Gamma_2
S_{\Omega_0}(y-x)\Gamma_1\bigr]
+{\rm Tr}\, \bigl[S_{\Omega_0}(0)\Gamma_2\bigr]\, 
{\rm Tr}\, \bigl[S_{\Omega_0}(0)\Gamma_1\bigr]\, .
\en
The above formula states that the closed fermion loop with two vertices
in the background of $\Omega_0$
is calculated replacing free nucleon propagators
by the in-medium fermion propagators 
$S_{\Omega_0}(x)$.
The same rule holds for a loop with any number of vertices, or any number of factorized loops.
Indeed, we may redefine the normal product with respect to the new vacuum
$|\Omega_0)$. Then, according to Eq.~(\ref{1prop}), the fermion 
propagator is just $S_{\Omega_0}(x)$. Since the combinatorial factors,
produced by the contractions in the Wick's theorem, 
do not depend on the explicit form of the 
propagator, one finds that all differences between closed 
fermion loops calculated in the vacuum and in the $\Omega_0$ background 
reduce to the
replacement of the nucleon propagator in the vacuum by $S_{\Omega_0}(x)$.
One concludes that, for a uniform fermion density distribution $\Omega_0$, the approach described in the present paper reduces
to the standard in-medium ChPT.

\subsection{Chiral power counting}
\label{sec:powercounting}

Another crucial ingredient for the approach proposed here is the
systematic bookkeeping based on chiral symmetry.
On first sight, the formalism allows too much freedom in the nuclear
matrix elements, which have to be evaluated for arbitrary space-time
arguments of the fields $\Psi(x)$. We shall now demonstrate that the
chiral power counting, in general, enables one to substantially reduce
this freedom. 

Even for a non-uniform distribution of  baryon number,  it is still
useful to think in terms of the 
``local Fermi momentum $k_F$'' which in this case is a function of the
space coordinate.  
It is common and convenient to count $k_F$ at $O(p)$ 
[the actual numerical value for this quantity in heavy nuclei, 
$k_F\sim 2m_{\pi}$, supports this conjecture]. Consequently, the
kinetic energies of the nucleons can be counted as $O(p^2)$. The
fermion field itself is $O(p^{3/2})$, $|\Omega)\sim O(1)$, and
the counting in the coefficient functions $S^{(n)}$ is standard.   
Furthermore, in order to be conform with the chiral power counting
for the pion   
self-energy operator (\ref{Pi}) in the uniform medium, we declare 
this quantity to be calculated at $O(p^k)$, if its actual power 
according to the above counting is $k-3$ (we remind the reader that,
for the uniform distribution, 
the self-energy operator reduces to $\Pi(E;{\bf q^\prime},{\bf q})\rightarrow
(2\pi)^3\delta^3({\bf q^\prime}-{\bf q})\,\tilde\Pi(E;{\bf q}^2)$, and $k$ thus
corresponds to the chiral power of $\tilde\Pi(E;{\bf q}^2)$, whereas
three additional powers are ``eaten'' by the $\delta$-function.).

The chiral power counting  described here implies a
systematic  expansion of the matrix elements in 
Taylor series, so that finally one ends up with the matrix elements with 
fields $\Psi(x)$ and derivatives thereof, taken at coinciding space-time 
arguments.  
Such matrix elements can in turn be written in terms of local 
distributions. This makes their construction transparent and reliable.

Below we shall schematically explain this method for the case of the
normal product of two fermion fields.
The corresponding contribution to the $\pi$-nucleus
scattering matrix element is given by
\eq\label{D1}
\bar D^{(1)}(q^\prime,q)=\int d^4\!x d^4\!y\,
D^{(1)}_{ba}(q^\prime,q|x,y)\,(\Omega|:\bar\Psi_b(x)
\Psi_a(y):|\Omega)\, ,
\en
where
\eq\label{one_insertion}
D^{(1)}_{ba}(q^\prime,q|x,y)=
\frac{-i}{(2BF)^2}\,\int d^4\!u d^4\!z\, {\rm e}^{iq^\prime \cdot u-iq\cdot z}
\frac{\delta^2 S^{(1)}_{ba}(j|x,y)}{\delta p^+(u)\delta p^-(z)}
\biggr|_{j=j_0}\, ,
\en
and $a=(\alpha,i),~b=(\beta,j)$ stand for multiindices with $\alpha,\beta$
and $i,j$ denoting Lorentz and isospin indices, respectively.
Furthermore, the threshold expansion is implied only for the quantity 
$S^{(1)}$ which is calculated in the vacuum. There is no need to perform the
 threshold 
expansion in the matrix element 
$(\Omega|:\bar\Psi_b(x)
\Psi_a(y):|\Omega)$, 
since, according to our picture of the nucleus,
all 3-momenta involved receive a cutoff at $k_F$.

Let us first consider the case of the standard HBChPT counting.
In this case, both energy and 3-momentum of the pions in the Feynman diagram
count like $O(p)$. In order to express the nuclear matrix elements in terms of
local distributions, we introduce the center-of-mass and
the relative coordinates $x=R+s/2$ and $y=R-s/2$. Since the nucleus is
static, the matrix element does not depend on the time component $R^0$
of $R$. Now, we consider
the dependence of this quantity on $s^0$. Factoring out the nucleon
(proton or neutron) mass  
$M_\Psi$,  
which is counted at $O(1)$, we can expand the rest in the Taylor series of 
$s^0$
\eq\label{s0}
&&(\Omega|:\bar\Psi_b(x)\Psi_a(y):|\Omega)=
{\rm e}^{iM_\Psi s^0}\,
(\Omega|:\bar N_b(0,{\bf x})
N_a(0,{\bf y}):|\Omega)
\nonumber\\[2mm]
&-&{\rm e}^{iM_\Psi s^0}\,\frac{s^0}{2}\,
(\Omega|:\bar N_b(0,{\bf x})\stackrel{\leftrightarrow}{\partial_0}
N_a(0,{\bf y}):|\Omega)+\cdots\, ,
\en
where $N_a(x)\doteq {\rm e}^{iM_\Psi x^0}\,\Psi_a(x)$.
Every time derivative acting on $N_a(x)$ produces
the kinetic energy of the nucleon, and has therefore to be counted as a 
quantity of  
order $p^2$. On the other hand, every power of $s^0$ corresponds to
the differentiation with respect to the energy variable in the Fourier
transform of $S^{(1)}$. 
Since, according to the standard HBChPT counting, differentiation of
every fermion propagator reduces the chiral power by $1$, one may count
$s^0$ at $O(p^{-1})$. We finally see that every term in the
expansion~(\ref{s0}) is $O(p)$ as compared to the previous term, so,
at a given order in the chiral expansion, one may truncate the series, and
arrives at a result that is local in the relative time variable
$s^0$.  

It is clear that this argumentation does not work for the spatial nonlocality:
here, every space derivative in the matrix element counts as $O(p)$, 
and spatial components $s^i\sim O(p^{-1})$, so one is not allowed to
truncate the Taylor series. One may, however, expand the fermion
propagators 
contained in $S^{(1)}$. In the HBChPT regime, this expansion reads 
\eq\label{HBChPT}
\frac{M_\Psi+\not\! p}{M_\Psi^2-p^2}=
\frac{1+\not\! v}{-2v\cdot l}+\frac{1}{M_\Psi}\,\biggl(
\frac{\not\! l}{-2 v\cdot l}
+\frac{(1+\not\! v)l^2}{4 (v\cdot l)^2}\biggr)+O(\frac{1}{M_\Psi^2})\, ,
\en
where $p=M_\Psi v+l$, the unit vector $v=(1,{\bf 0})$, and as usual
${\not p}=p_\mu \gamma^{\mu}$. 
From this expansion one observes that i) each term with an additional power
of $M_\Psi^{-1}$ is $O(p)$
suppressed with respect to the previous term, so the series can be
truncated at a given chiral order, and ii) each term is a polynomial
in $3$-momenta. According to the above properties, at a given order in
the chiral expansion, $S^{(1)}$ in coordinate space is a finite sum of
spatial $\delta$-functions and derivatives thereof. 
Substituting into Eq.~(\ref{D1}) and integrating by parts, one finally 
obtains a finite sum containing matrix elements of the fermion
bilinears and derivatives thereof, taken at the same space-time point
$R$. In addition, we recall that the field $\Psi$ obeys the free Dirac
equation, so all time derivatives can be eliminated at the end. 

The expression which one obtains after performing all Taylor
expansions, is much 
simpler than the original one. Whereas  the matrix element in
Eq.~(\ref{D1}) involves a ``dynamical'' density matrix (with
non-coinciding time variables), which is strongly dependent on the
description of the structure of the nucleus, after
the Taylor expansion one gets terms of the type 
\eq\label{local}
(\Omega|:\bar\Psi_b(0,{\bf R})
\stackrel{\leftrightarrow}{\partial_{i_1}}
\cdots\stackrel{\leftrightarrow}{\partial_{i_n}}
\Psi_a(0,{\bf R}):|\Omega)\, ,\quad\quad
n=0,1,\cdots\, ,
\en
which have a simpler interpretation. For example, using the Fierz
transformation, the term without derivatives can be related to the
matrix elements of local scalar, vector, $\ldots$ fermion bilinears 
  $(\Omega|:\bar\Psi\Gamma_i\Psi:|\Omega)$
with $\Gamma_i=1,\gamma_\mu,\cdots$. To lowest order in the chiral expansion, these matrix elements coincide with the scalar, vector 
$\ldots$ formfactors of the nucleus, which can be parametrized, e.g.,
in terms of the pertinent radii, without any more detailed explicit
knowledge of the  nuclear structure.

Next, we briefly consider the case of the non-standard counting.
As it will be demonstrated below by explicit calculations, 
the contributions to the self-energy, corresponding to the non-standard 
counting regime, do not appear up to and including $O(p^5)$ in the pion 
two-point function in the vicinity of the mass shell. For the general pion
kinematics this is not always the case. In particular, it has been shown
\cite{Leutwyler} that, for a large class of the collective phenomena,
the elementary excitations can be described by the non-relativistic
dispersion law $E=\gamma{\bf p}^2$. Moreover, it has been argued, that the 
same dispersion law applies in the Goldstone boson condensed phase of the 
nuclear ground state (see e.g. \cite{Toublan} where this issue is considered
in detail). So, in order to be able to use \NChPT to study the phenomenon of the
pion (and kaon) condensation in general \cite{EW,Kaplan,Politzer,others}, 
it is necessary to modify counting rules for the external momenta and to assign 
$E\sim p^2$ to the energy, while 3-momentum counting remains standard 
\cite{Annals}. At the next step, one has to look for the pole in the 
two-point function with the vanishing energy $E$, which signals that the 
condensation has indeed occurred. One sees (section 4.2 of Ref. \cite{Annals})
that in the asymmetric matter where the density exceeds the critical value 
given by Eq. (4.10) of this paper, 
the inverse pion propagator indeed develops a pole
at $E,{\bf p}^2\to 0$ and $D={\bf p}^2/E$ finite. This means that for the
densities that obey the condition (4.10), the non-standard counting may 
appear already at the tree level. Note as well, that Eq. (4.10) implies
that the density is counted at $O(p^2)$ rather that at $O(p^3)$. 
Another example, where the non-standard counting appears at the leading
order, is the in-medium $\pi\pi$ scattering amplitude \cite{Annals}.
In this paper, we do not consider the non-standard regime in detail since
in the context of pionic atoms the contribution from this regime arises
at the one-loop level. In this case, our approach should be adapted 
correspondingly.

To summarize, chiral power counting helps to systematically organize
the perturbation expansion in the background of a static nucleus, in
terms of  {\em local} distributions. This construction, however,
heavily relies on the assumption that only the momentum regime
relevant in HBChPT,  gives rise to non-zero contributions in the 
coefficient functions  $S^{(n)}$.  This conjecture may or may not hold 
-- in the latter case one has
the non-standard counting which has to be dealt with separately. Non-local
quantities, describing more detailed aspects of  nuclear structure, may
become necessary here. 

In the section~\ref{sec:Pi} we present the explicit calculation of the
pion self-energy operator at $O(p^5)$. As already mentioned above,
we demonstrate that only the
HBChPT momentum regime is relevant in Feynman integrals at this order,
so the answer can be given entirely in terms of local distributions. 
In this paper we do not attempt to analyze higher orders in the chiral 
expansion.

\subsection{The static limit}
\label{sec:recoil}

In this section we examine in some detail the reasoning behind
 the picture of a static nucleus, and explain the meaning of the 
``round-bracket'' notation which was introduced in Eq. (\ref{def-D1}).
 As was stated, the procedure involves two steps when
 evaluating the matrix elements~(\ref{def-D}) in the background of a
 (heavy) nucleus: first, removing a momentum conserving delta function
 in the CM frame [which coincides with the nuclear rest frame for an
 infinitely heavy nucleus], and secondly, passing to the static
 nucleus limit in terms of the {\em velocities} 
of the initial and final nuclei.

For the sake of illustration, let us first consider the term in the two-point
function (\ref{def-D}), corresponding
to the in-vacuum self-energy of the pion:
\eq\label{Z0-bb}
\hspace*{-.4cm}&&\frac{-i}{(2BF)^2}\,\int d^4\!x d^4\!y \,{\rm
e}^{iq^\prime \cdot x-iq\cdot y}
\langle\Omega ({\bf P'})|\frac{\delta^2 S^{(0)}(j)}{\delta p^+(x)\delta p^-(y)}
|\Omega ({\bf P})\rangle\biggr|_{j=j_0}
\nonumber\\[2mm]
\hspace*{-.4cm}&=&(2\pi)^3\delta^3({\bf P'}+{\bf q^\prime}-{\bf P}-{\bf q})\,
\frac{-i}{(2BF)^2}\,\int d^4\!x d^4\!y \,{\rm e}^{i q^\prime \cdot 
x-iq\cdot y}
(\Omega|
\frac{\delta^2 S^{(0)}(j)}{\delta p^+(x)\delta p^-(y)}
|\Omega)\biggr|_{j=j_0} ,
\nonumber\\
\hspace*{-.4cm}&&
\en
with $(\Omega|\Omega)=1$.

Next, consider the term with one medium insertion. Using translational
invariance and taking the limit $M_\Omega\to\infty$,
we obtain [cf Eq.~(\ref{D1})] in the CM frame:
\eq\label{Z1-bb}
&&\int d^4\!x d^4\!y  D^{(1)}_{ba}(q^\prime,q|x,y)
\langle\Omega ({\bf P'})|:\bar\Psi_b(x)\Psi_a(y):|\Omega ({\bf P})\rangle
\nonumber\\[2mm]
&\!\!\!=\!\!\!&
(2\pi)^3\delta^3({\bf P'}+{\bf q^\prime}-{\bf P}-{\bf q})
\int d^4\!x d^4\!y D^{(1)}_{ba}(q^\prime,q|x,y)
(\Omega|:\bar\Psi_b(x)\Psi_a(y):|\Omega)\, ,
\nonumber\\
&&
\en
where
\eq\label{sachs}
&&\hspace*{-.8cm}(\Omega|:\bar\Psi_b(x)\Psi_a(y):|\Omega)=
\int\frac{d^3{\bf Q}}{(2\pi)^3}\,{\rm e}^{i{\bf Q}\cdot  \frac{{\bf x}+{\bf y}}{2}}\,
f_{ba}({\bf Q},{\bf x}-{\bf y})\, ,
\nonumber\\[2mm]
&&\hspace*{-.8cm}f_{ba}({\bf Q},{\bf x}-{\bf y})
=\langle\Omega(-{\bf L}-\frac{{\bf Q}}{2})|
:\bar\Psi_b\biggl(0,\frac{{\bf x}-{\bf y}}{2}\biggr)
\Psi_a\biggl(0,-\frac{{\bf x}-{\bf y}}{2}\biggr):|\Omega(-{\bf L}+\frac{{\bf Q}}{2})\rangle\, ,\nonumber\\
&&
\en
and ${\bf L}=\frac{1}{2}\, ({\bf q^\prime}+{\bf q})$ [Note that in the static
limit $M_\Omega\to\infty$ 
the bracket matrix element in Eq.~(\ref{sachs}) should not depend on
the momentum  
${\bf L}$, but only on the momentum transfer ${\bf Q}$ since, in this
limit, the  
round-bracket matrix element in the same equation should depend on the
vectors ${\bf x},{\bf y}$ alone.]. 
From these examples we arrive at the following interpretation.
The round-bracket state $|\Omega)$ denotes the static nucleus
fixed at a given point of space (at the origin).
As  mentioned already, this state is assumed to be spin-saturated for
simplicity. 
The expectation value of any operator in the state
$|\Omega)$ is translation-invariant in time (but not in
space).  
Furthermore, this expectation value (round-bracket matrix element) is
equal to the Fourier  
transform of a corresponding formfactor evaluated in the Fock space.
The generalization for the higher-order terms is straightforward.

Next, we consider in more detail the role of the restriction to
spin-saturated  states $\Omega$. Let us start with the matrix element
of the operator 
$:\bar\Psi(x)\gamma_\mu\Psi(x):$. 
In order to pass easily to the static limit, it is advantageous to
work in terms of velocities
$v_\mu=P_\mu/M_\Omega$, ${v'}_\mu={P'}_\mu/M_\Omega$
instead of momenta $P_\mu,~{P'}_\mu$.  
The above current is conserved since $\Psi(x)$ is a free field.  Then,
from the assumption that 
the bound state of the nucleus does not depend on the internal quantum
numbers, it follows that the Fock space matrix element of this
current is described by a single scalar formfactor: 
\eq\label{heavymass} 
\langle\Omega({\bf P'})|:\bar\Psi(0)\gamma_\mu\Psi(0):|\Omega ({\bf P})\rangle
=\frac{1}{2}\, (v'+v)_\mu F(t)\, , \quad\quad t=(P'-P)^2\, ,
\en
where $F(t)$ is the form factor of the charge distribution related to
the current $\bar \Psi \gamma_\mu \Psi$.
In the static limit, when $v_\mu,{v'}_\mu\to (1,{\bf 0})$ and
$t\to -({\bf P'}-{\bf P})^2=-({\bf q^\prime}-{\bf q})^2$, according to
Eq.~(\ref{sachs}), this matrix element reduces to
\eq\label{heavymass_static}
(\Omega|:\bar\Psi(x)\gamma_\mu\Psi(x):|\Omega)=
g_{\mu 0}\int \frac{d^3{\bf Q}}{(2\pi)^3}\,{\rm e}^{i{\bf Q} \cdot {\bf x}}\, F(-{\bf Q}^2)\, .
\en
Other fermion bilinears can be considered analogously. For example,
symmetry considerations imply that  the matrix element of the operator
$:\bar\Psi(x)\sigma_{\mu\nu}\Psi(x):$ can involve
 only the Lorentz-structure 
${v'}_\mu v_\nu-{v'}_\nu v_\mu$. Then, in the static limit, the
round-bracket matrix element should vanish: 
\eq\label{sigma_mn}
(\Omega|:\bar\Psi(x)\sigma_{\mu\nu}\Psi(x):|\Omega)=0\, .
\en
To summarize, the use of equations of motion and symmetries, together
with the simplifying choice of a  spin-saturated 
background nucleus 
$\Omega$, greatly reduces the number of independent round-bracket
matrix elements which serve as an empirical input 
in the construction of the self-energy operator.

Finally, we mention that for the free Fermi-gas the equality 
$(\Omega_0|\Omega_0)=1$ and further, the
equations~(\ref{commutators}) in the rigorous sense are understood as 
follows
\eq\label{sense}
&&\langle\Omega_0|b^\dagger({\bf l}_1,s_1)b({\bf l}_2,s_2)|\Omega_0\rangle
=(2\pi)^32l_1^0\delta^3({\bf l}_1-{\bf l}_2)\theta(k_F-|{\bf l}_1|)
\langle\Omega_0|\Omega_0\rangle\, ,
\en
and so forth.

\setcounter{equation}{0}
\section{Pion self-energy at $O(p^5)$}
\label{sec:Pi}

In this section we demonstrate the general rules formulated in
previous sections, by
presenting the detailed calculation of the pion self-energy 
operator at $O(p^5)$ in ChPT in the presence of the finite nucleus.
This is an instructive exercise, even though the calculation up to
this order yields just the leading terms of the pion-nuclear optical
potential, those linear in proton and neutron densities.

\subsection{Pion self-energy in the vacuum}

\begin{figure}[t]
\begin{center}
\includegraphics[width=13.cm]{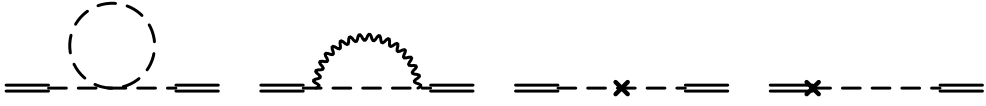}
\end{center}
\caption{
Diagrams contributing to the two-point function of the pseudoscalar densities
at $O(p^4)$ in the $\sigma$-model parameterization. Double, dashed and wiggle
lines stand for the pseudoscalar densities, pions and photons, respectively.
Crosses denote the vertices with $O(p^4)$ LECs $l_i,k_i$.
}
\label{fig:invacuum}
\end{figure}

The two-point function of the pseudoscalar densities in the vacuum is given by [cf with Eqs.~(\ref{SZ}) and (\ref{def-D})]
\eq\label{D0}
&&\hspace*{-.7cm}
(2\pi)^4\delta^4(q^\prime-q)D^{(0)}(q^2)=
\frac{-i}{(2BF)^2}\int d^4\!x d^4\!y \, {\rm e}^{iq^\prime\cdot x-iq\cdot y}
\frac{\delta^2 S^{(0)}(j)}{\delta p^+(x)\delta p^-(y)}
\biggr|_{j=j_0} .
\en
Up to and including $O(p^4)$, only the diagrams shown in Fig.~\ref{fig:invacuum} contribute to the self-energy of the pseudoscalar densities in the vacuum. 
The calculations are most easily done in the so-called $\sigma$-model 
parameterization
\eq\label{sigmamodel}
U=\biggl(1-\frac{\mathbold{\pi}^2}{F^2}\biggr)^{1/2}+
\frac{i\mathbold{\tau \cdot \pi}}{F}\, ,
\en
where $\mathbold{\pi}(x)$ denotes the interpolating pion field.
We use this parameterization throughout this paper [of course, the two-point 
function of the pseudoscalar densities does not depend on the choice of a
particular parameterization].
A straightforward calculation with the use of the Lagrangian from 
appendix~\ref{app:lagr} gives
\eq\label{D0p2}
D^{(0)}(q^2)=\frac{1}{q^2 - m_{\pi}^2}+\frac{1}{q^2 - m_{\pi}^2}
\tilde \Pi^{(0)}(q^2)
\frac{1}{q^2 - m_{\pi}^2}+\cdots\, ,
\en
where the dots stand for higher-order terms in the chiral expansion, and
\eq\label{Mplus}
\tilde\Pi^{(0)}(q^2)&=&
(q^2 - m_{\pi}^2)A^{(0)}+(q^2 - m_{\pi}^2)^2B^{(0)}(q^2)\, ,
\en
where $B^{(0)}(q^2)$ is finite at $q^2\to m_{\pi}^2$ and $d\neq 4$.
The pertinent contribution to the self-energy is given by
\eq\label{Pi0}
\Pi^{(0)}(E;{\bf q^\prime},{\bf q})=(2\pi)^3\delta^3({\bf q^\prime}-{\bf q})
\tilde\Pi^{(0)}(E^2-{\bf q}^2)\, .
\en
The expression for $A^{(0)}$ takes the form
\eq\label{A0}
A^{(0)}&=&\frac{m_{\pi}^2}{F^2}\,\biggl(2l_4^r-\frac{1}{16\pi^2}\,
\ln\frac{m_{\pi}^2}{\mu^2}\biggr)+\frac{4m_{\pi^0}^2}{F^2}\, l_3^r
+e^2[2(3-\xi)\lambda_{\rm IR}
\nonumber\\[2mm]
&-&\frac{20}{9}\,(k_1^r+k_2^r-2k_5^r-2k_6^r)+\frac{8}{9}\,k_7^r+8k_8^r]
+O(d-4)\, ,
\en
where $\mu$ denotes the scale of the dimensional regularization. 
Note that we tame both ultraviolet and infrared divergences in
this  regularization, and we attach the subscript "IR" to
\eq\label{lambdaIR}
\lambda_{\rm IR}=\frac{\mu^{d-4}}{16\pi^2}\,
\biggl(\frac{1}{d-4}-\frac{1}{2}\,(\Gamma'(1)+\ln 4\pi+1)\biggr)\, ,
\en
in order to distinguish ultraviolet and infrared divergences.
Finally, we note that the explicit expression for $B^{(0)}(p^2)$   is
never needed, because it does not contribute to the residue of the
two-point function at this order of the chiral expansion -- for this
reason we do not display it here.  

As one immediately sees from Eq.~(\ref{A0}), the pion self-energy is both infrared-divergent and gauge-dependent off-shell, i.e. when 
$q^2\neq m_{\pi}^2$. 
Below we shall demonstrate that in the eigenvalue equation for
determining the binding energy of the pion-nuclear bound state, where
pion self-energy plays the role of the potential, such off-shell effects
begin to contribute only at higher chiral order and can be neglected
at the level  we are working.  

\subsection{One medium insertion}

\begin{figure}[t]
\begin{center}
\includegraphics[width=13.cm]{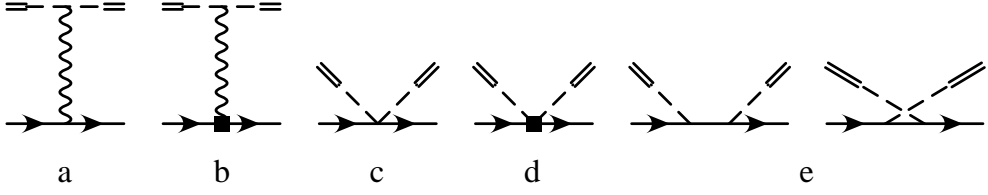}
\end{center}
\caption{
Diagrams contributing to the quantity $D^{(1)}$ up to and including $O(p^5)$:
(a) one-photon exchange;
(b) contributions from the anomalous magnetic moment of the nucleon;
(c) Weinberg-Tomozawa vertex;
(d) contributions from LECs $c_i,f_i$;
(e) nucleon pole diagrams.
}
\label{fig:D1}
\end{figure}

According to Eq.~(\ref{SFock}), the contribution with one medium 
insertion is expressed in terms of the coefficient function 
(\ref{one_insertion}) where, up to and including $O(p^5)$, the quantity 
$D^{(1)}$ is given by the
diagrams in Fig.~\ref{fig:D1}. These are: the one-photon exchange diagram, 
contributions from the anomalous magnetic moment of the nucleon,
the Weinberg-Tomozawa vertex, contributions from the 
LECs $c_i,f_i$, and the nucleon pole diagrams
\eq\label{D1_sum}
&&({q^\prime}^2-m_{\pi}^2)(q^2-m_{\pi}^2)D^{(1)}_{ba}(q^\prime,q|x,y)
=\delta^4(x-y){\rm e}^{i(q^\prime-q)\cdot x}
[D^{(1\gamma)}_{ba}(q^\prime,q)
\nonumber\\[2mm]
&&+\,D^{(an)}_{ba}(q^\prime,q)
+D^{(WT)}_{ba}(q^\prime,q)
+D^{(cf)}_{ba}(q^\prime,q)]
+D^{(P)}_{ba}(q^\prime,q|x,y)\, ,
\en
where we use the physical value of the squared pion mass, $m_{\pi}^2$,
instead of its $O(p^2)$ value, since the difference is of higher order
in ChPT, which we neglect anyway. Further, 
\eq\label{DDDD}
D^{(1\gamma)}_{ba}(q^\prime,q)&=&e^2(\gamma_\mu)_{\beta\alpha}
\frac{1}{2}\,(1+\tau^3)_{ji}(q^\prime+q)_\nu D^{\mu\nu}(q^\prime-q)\, ,
\nonumber\\[2mm]
D^{(an)}_{ba}(q^\prime,q)&=&\frac{e^2}{m}(i\sigma_{\mu\lambda})_{\beta\alpha}
\biggl(\frac{c_6+2c_7}{4}+\frac{c_6}{4}\,\tau^3\biggr)_{ji}
(q^\prime+q)_\nu (q-q^\prime)^\lambda D^{\mu\nu}(q^\prime-q) ,
\nonumber\\[2mm]
D^{(WT)}_{ba}(q^\prime,q)&=&-\frac{1}{4F^2}\,(\gamma_\mu)_{\beta\alpha}
(\tau^3)_{ji} (q^\prime+q)^\mu\, ,
\nonumber\\[2mm]
D^{(cf)}_{ba}(q^\prime,q)&=&\biggl(\frac{4c_1m_{\pi^0}^2-2c_2{q^0}^\prime
q^0 -2c_3(q^\prime q)}{F^2}+2e^2f_1\biggr)\delta_{\beta\alpha}\delta_{ji}
\nonumber\\[2mm]
&+&\frac{e^2}{2}\,f_2\delta_{\beta\alpha}(\tau^3)_{ji}
+\frac{ic_4}{F^2}\,q^\prime_\nu q_\mu \sigma^{\mu\nu}_{\beta\alpha}\tau^3_{ji}\, ,
\nonumber\\[2mm]
D^{(P)}_{ba}(q^\prime,q|x,y)&=&-\frac{g_A^2}{2F^2}\,
(\not\! q^\prime \gamma_5S_n(x-y)\not\! q\gamma_5)_{\beta\alpha}
\frac{1}{2}(1+\tau^3)_{ji}{\rm e}^{iq^\prime \cdot x-iq\cdot y}
\nonumber\\[2mm]
&-&\frac{g_A^2}{2F^2}\,
(\not\! q\gamma_5S_p(x-y)\not\! q^\prime \gamma_5)_{\beta\alpha}
\frac{1}{2}(1-\tau^3)_{ji}{\rm e}^{iq^\prime \cdot y-iq\cdot x}\, .
\en
In these formulae, $D^{\mu\nu}$ and $S_{p,n}$ stand for the free
photon and baryon propagators, respectively:
\eq\label{free_prop}
D^{\mu\nu}(x)&=&\int\frac{d^4l}{(2\pi)^4}\,{\rm e}^{-il\cdot x}\,\frac{1}{l^2}\,
\biggl( g^{\mu\nu}-(1-\xi)\frac{l^\mu l^\nu}{l^2}\biggr)\, ,
\nonumber\\[2mm]
S_i(x)&=&\int\frac{d^4l}{(2\pi)^4}\,{\rm e}^{-il\cdot x}\,
\frac{1}{M_i-\not l}\, .
\en
The corresponding contribution to the self-energy is
\eq\label{Pi1}
2\pi\delta({q^0}^\prime -q^0)\Pi^{(1)}(E;{\bf q^\prime},{\bf q})=\int
d^4\!x d^4\!y
D^{(1)}_{ba}(q^\prime,q|x,y)(\Omega|:\bar\Psi_b(x)\Psi_a(y):|\Omega 
)\, .
\en
We proceed to investigate term by term in detail
the contributions~(\ref{DDDD}) to the two-point function of the pseudoscalar
densities.

\subsubsection{Coulomb potential and contribution from the anomalous magnetic moment of the nucleon}

According to Eq.~(\ref{DDDD}), the one photon exchange contribution to the 
two-point function of the pseudoscalar densities is given by
\eq\label{D1gamma_SE}
\bar D^{(1\gamma)}(q^\prime,q)=\frac{1}{{q^\prime}^2-m_{\pi}^2}\,
2\pi\delta({q^0}^\prime-q^0)\Pi^{(1\gamma)}(E;{\bf q^\prime},{\bf
q})\frac{1}{q^2-m_{\pi}^2} 
\, ,
\en
where
\eq\label{P1gamma_SE}
2\pi\delta({q^0}^\prime-q^0)\Pi^{(1\gamma)}(E;{\bf q^\prime},{\bf q})
&=&e^2\int d^4\!x{\rm e}^{i(q^\prime-q)\cdot x}(q^\prime+q)_\nu D^{\mu\nu}(q^\prime-q)\times
\nonumber\\[2mm]
&\times&
(\Omega|:\bar\Psi(x)\gamma_\mu\frac{1}{2}\,(1+\tau^3)\Psi(x):|
\Omega)\, .
\en
In order to demonstrate that the longitudinal piece 
depending  on the gauge parameter
$\xi$ does not contribute in Eq.~(\ref{P1gamma_SE}), we write this
matrix element in the form 
\eq\label{charge_density}
(\Omega|:\bar\Psi(x)\gamma_\mu\frac{1}{2}\,(1+\tau^3)\Psi(x):|
\Omega)=g_{\mu0}\frac{1}{2}\,(\rho^0({\bf x})+\rho^3({\bf x}))\, ,
\en
with the isoscalar and isovector densities
$\rho^0 = \rho_p + \rho_n$ and $\rho^3 = \rho_p - \rho_n$. The density
in (\ref{charge_density}) is related to the Fourier-transform of the
formfactor of the nuclear charge distribution,
\eq\label{density_Fourier}
\frac{1}{2}\,(\rho^0({\bf x})+\rho^3({\bf x}))= \rho_p({\bf x}) =
\int\frac{d^3{\bf l}}{(2\pi)^3}\,{\rm e}^{i{\bf l}\cdot {\bf x}}
F_{0+3}[-{\bf l}^2]\, ,
\en
where
\eq\label{0+3}
\langle\Omega'({\bf P^\prime})|:\bar\Psi(0)\gamma_\mu\frac{1}{2}\,(1+\tau^3)\Psi(0):|
\Omega ( {\bf P}) \rangle=\frac{1}{2}\, (v'+v)_\mu F_{0+3}[t]
\en
with
$t=(P_\Omega'-P_\Omega)^2$ [cf Eqs.~(\ref{heavymass})
(\ref{heavymass_static})]. At the order in ChPT we are working,
the formfactor $F_{0+3}[t]$ is proportional to the 
electromagnetic formfactor of the nucleus $\Omega$

The presence of $\delta(p^0-q^0)$ in Eq.~(\ref{D1gamma_SE})
 implies that  the one-photon
exchange contribution turns out to be effectively instantaneous.
Eq.~(\ref{P1gamma_SE}) takes the form
\eq\label{P1gamma_r}
\Pi^{(1\gamma)}(E;{\bf q^\prime},{\bf q})=-2E\int d^3{\bf x}
{\rm e}^{-i({\bf q^\prime}-{\bf q})\cdot {\bf x}}V_C({\bf x})\, ,
\en
where
\eq\label{VC}
V_C({\bf x})=\int d^3{\bf r}\,\frac{\alpha}{|{\bf x}-{\bf r}|}\,
\rho_p({\bf r})\, ,
\en
and $\alpha=e^2/(4\pi)$ denotes the fine structure constant.
Recall that, in the chiral counting, $\rho^{[0,3]}({\bf r})\sim p^3$ and $e\sim p$.
It is straightforward to observe that the quantity $\Pi^{(1\gamma)}$ given
by Eq.~(\ref{P1gamma_r}), counts as $O(p)$. According to our conventions
from section~\ref{sec:framework} [the actual chiral power of 
$\Pi(E;{\bf q^\prime},{\bf q})$ differs by 3 units from the chiral power assigned, 
in order to stay in conformity with the chiral counting in the vacuum], 
the Coulomb term contributes to the self-energy at $O(p^4)$.

We finally note that the contribution from the anomalous magnetic moment of the
nucleon -- which otherwise would contribute at $O(p^5)$ --
vanishes since the 
matrix element of the operator $:\bar\Psi\sigma_{\mu\nu}\Psi:$ between the
nuclear states disappears in the static limit [see Eq.~(\ref{sigma_mn})]. Consequently,
\eq\label{Pi_an}
\Pi^{(an)}(E;{\bf q^\prime},{\bf q})=0\, .
\en

\subsubsection{Contact contributions}

Contact contributions include: the Weinberg-Tomozawa term [$O(p^4)$], and
the LEC contributions [$O(p^5)$] involving $c_i$ and $f_i$ . From
Eq.~(\ref{DDDD}) one obtains 
\eq\label{contact}
&&\hspace*{-.7cm}\Pi^{(WT)+(cf)}(E;{\bf q^\prime},{\bf q})
=\int d^3{\bf x}{\rm e}^{-i({\bf q^\prime}-{\bf q}) \cdot {\bf x}}\,[\Pi_0(E;{\bf x})
+{\bf q^\prime}\cdot {\bf q}\Pi_2(E;{\bf x})]
\, ,
\nonumber\\[2mm]
&&\hspace*{-.7cm}\Pi_0(E;{\bf x})
=-\frac{E}{2F^2}\,\rho^3({\bf x})
+\biggl(\frac{4c_1m_{\pi^0}^2-2(c_2+c_3)E^2}{F^2}+2e^2f_1\biggr)
\sigma^0({\bf x})
+\frac{e^2}{2}\, f_2\sigma^3({\bf x})\, ,
\nonumber\\[2mm]
&&\hspace*{-.7cm}\Pi_2(E;{\bf x})=\frac{2c_3}{F^2}\,\sigma^0({\bf x})\, ,
\en
with the isoscalar and isovector scalar densities
\eq\label{sigma-03}
\sigma^{[0,3]}({\bf x})=(\Omega|
:\bar\Psi(x)[1,\tau^3]\Psi(x):|\Omega)\, .
\en

Note that the contribution proportional to $c_4$, vanishes for the static 
nucleus. 
We would like to stress that, in order to be consistent with the chiral 
counting in the nuclear matrix elements and our description of the nucleus, 
one has to count the difference
between $\sigma^{[0,3]}({\bf x})$ and $\rho^{[0,3]}({\bf x})$ at a higher
chiral order than these quantities themselves. 
This difference, which involves 
 $1-\gamma_0$, is determined by the overlap of the ``small'' components
of the Dirac wave function of the nucleon. For a slowly moving nucleon
inside the nucleus, this difference is suppressed by a factor $p^2$ as compared to the overlap 
of the ``large'' components. This can be directly seen for the case of the
uniform density, where $\rho^{[0,3]}-\sigma^{[0,3]}\sim O(k_F^5)$, whereas 
each term individually is of order $k_F^3$. Below, we shall always use the
counting
\eq\label{rho-sigma}
\frac{\rho^{[0,3]}({\bf x})-\sigma^{[0,3]}({\bf x})}{\rho^{[0,3]}({\bf x})}
\sim O(p^2)\, ,
\en
and eliminate $\sigma^{[0,3]}({\bf x})$ in favor of 
$\rho^{[0,3]}({\bf x})$ in all expressions.

Note also that one may relate the matrix elements (\ref{sigma-03}) to the
scalar formfactor of the nucleus
\eq\label{s-03}
\sigma^0({\bf x})&=&\frac{1}{8ic_1B}\,\frac{\delta}{\delta s^0(x)}\,
(\Omega|\hat S(j)|\Omega)\biggr|
_{j=j_0}+\cdots\, ,
\nonumber\\[2mm]
\sigma^3({\bf x})&=&\frac{1}{4ic_5B}\,\frac{\delta}{\delta s^3(x)}\,
(\Omega|\hat S(j)|\Omega)\biggr|
_{j=j_0}+\cdots\, ,
\en
modulo higher-order terms in chiral expansion.
At the order we are working, these higher-order terms can be neglected.

\subsubsection{Nucleon loop in the presence of the nucleus}

The contribution coming from the diagrams of 
Fig.~\ref{fig:D1}e, involves
the matrix element of two fermion fields taken at different space-time points.
In order to evaluate this contribution, one has to use the expansion 
described in sections~\ref{sec:powercounting}, \ref{sec:recoil}, along with 
the bookkeeping based on chiral symmetry. A straightforward, yet 
tedious calculation leads to:
\eq\label{Pi-P}
\Pi^{(P)}(E;{\bf q^\prime},{\bf q})=\frac{g_A^2}{2F^2}\int d^3{\bf x}
{\rm e}^{-i({\bf q^\prime}-{\bf q}) \cdot {\bf x}} [\Pi_4(E;{\bf
q^\prime},{\bf q}|{\bf x})+ 
\Pi_5(E;{\bf q^\prime},{\bf q}|{\bf x})]+\cdots\, ,
\en
where
\eq\label{pi_45}
\Pi_4(E;{\bf q^\prime},{\bf q}|{\bf x})&\!\!=\!\!&\,\frac{{\bf
q^\prime}\cdot {\bf q}}{E}\, 
\rho^3({\bf x}) ,
\nonumber\\[2mm]
\Pi_5(E;{\bf q^\prime},{\bf q}|{\bf x})&\!\!=\!\!&
\frac{(M_n-M_p){\bf q^\prime} \cdot {\bf q}}{E^2}\,\rho^3({\bf x})
\nonumber\\[2mm]
&+&\biggl(\frac{E^2-2{\bf q^\prime}\cdot {\bf q}}{2M_N}
+\frac{3({\bf q^\prime}\cdot {\bf q})^2-{\bf q^\prime}^2{\bf q}^2}{4M_NE^2}\biggr)
\rho^0({\bf x})\, .
\en
Here $M_N=\frac{1}{2}\,(M_p+M_n)$, and the dots
 in Eq.~(\ref{Pi-P}) stand
for the higher-order terms in chiral expansion.

\subsection{Two medium insertions}

\begin{figure}[t]
\begin{center}
\includegraphics[width=8.cm]{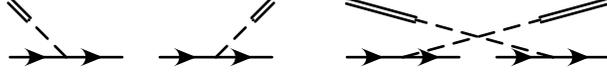}
\end{center}
\caption{
Two medium insertions: the
diagrams contributing to the quantity $D^{(2)}$, Eq.~(\ref{D2-dd}).
}
\label{fig:2-medium}
\end{figure}

The coefficient function $D^{(2)}$, which corresponds to the
two medium insertions in the pion self-energy, is given by
\eq\label{two_insertions}
&&D^{(2)}_{b_1b_2a_1a_2}(q^\prime,q|x_1,x_2,y_1,y_2)
\nonumber\\[2mm]
&=&
\frac{-i}{(2BF)^2}\,\int d^4\!u d^4\!z\, {\rm e}^{iq^\prime \cdot u-iq\cdot z}
\frac{\delta^2 S^{(2)}_{b_1b_2a_1a_2}(j|x_1,x_2,y_1,y_1)}
{\delta p^+(u)\delta p^-(z)}
\biggr|_{j=j_0}\, .
\en
Up to and including $O(p^5)$, there is a single contribution which stems from 
the pseudovector pion-nucleon Lagrangian. 
This contribution, which is 
diagrammatically shown in Fig~\ref{fig:2-medium}, equals
\eq\label{D2-dd}
\hspace*{-.5cm}&&
({q^\prime}^2-m_{\pi}^2)(q^2-m_{\pi}^2)D^{(2)}_{b_1b_2a_1a_2}(q^\prime,q|x_1,x_2,y_1,y_2)=  
\frac{ig_A^2}{16F^2}\,
{\rm e}^{iq^\prime \cdot x_1-iq\cdot x_2}\times
\nonumber\\[2mm]
\hspace*{-.5cm}&\times&
\delta^4(x_1-y_1)\delta^4(x_2-y_2)
(\not\! q^\prime \gamma_5)_{\beta_1\alpha_1}(\tau^1+i\tau^2)_{j_1i_1}
(\not\! q\gamma_5)_{\beta_2\alpha_2}(\tau^1-i\tau^2)_{j_2i_2}
\nonumber\\[2mm]
&+&crossed~term\, .
\en

The pertinent contribution to the self-energy is
\eq\label{4fermion}
&&2\pi\delta({q^0}^\prime-q^0)\Pi^{(2)}(E;{\bf q^\prime},{\bf
q})=-\frac{ig_A^2}{16F^2}\, 
\int d^4\!x d^4\!y \,{\rm e}^{iq^\prime \cdot x-iq\cdot y}\times
\nonumber\\[2mm]
&\times& \!\! (\Omega|
:\bar\Psi(x)\not\! q^\prime \gamma_5(\tau^1+i\tau^2)\Psi(x)
\bar\Psi(y)\not\! q\gamma_5(\tau^1-i\tau^2)\Psi(y):|\Omega)+crossed~term\, .
\nonumber\\
\en
It is convenient to work in the momentum representation for the free fields
$\Psi$. For the self-energy, one gets the following expression
\eq\label{momentum_rep}
\hspace*{-.4cm}&&2\pi\delta({q^0}^\prime -q^0)\Pi^{(2)}(E;{\bf
q^\prime},{\bf q})=\frac{ig_A^2}{16F^2}\, 
\sum_{s_1s_2s_3s_4}\int\frac{d^4l_1}{(2\pi)^4}\,\frac{d^4l_2}{(2\pi)^4}\, 
\theta(l_1^0)\delta(l_1^2-M_p^2)\times
\nonumber\\[2mm]
\hspace*{-.4cm}&\times&\theta(l_1^0+{q^0}^\prime)\delta((l_1+q^\prime)^2-M_n^2)
\theta(l_2^0)\delta(l_2^2-M_n^2)\theta(l_2^0-q^0)\delta((l_2-q)^2-M_p^2)\times
\nonumber\\[2mm]
\hspace*{-.4cm}&\times&\bar u({\bf l}_1,s_1)\not\! q^\prime \gamma_5u({\bf l}_1+q^\prime,s_3)\,\,
\bar u({\bf l}_2,s_2)\not\! q\gamma_5u({\bf l}_2-q,s_4)\times
\nonumber\\[2mm]
\hspace*{-.4cm}&\times&
(\Omega|b^\dagger({\bf l}_1s_1)b^\dagger({\bf l}_2s_2)b({\bf l}_3s_3)b({\bf l}_4s_4)|
\Omega)+\cdots\, ,
\en
where $u({\bf l}s)=\pmatrix{u_p({\bf l}s)\cr u_n({\bf l}s)}$ denotes the 
8-component Dirac spinor of the nucleon, and the dots indicate terms with 
creation/annihilation operators for 
 antinucleons, and the crossed term.
Below, we shall demonstrate that, in analogy with the infinite nuclear 
medium~\cite{Annals}, the above contribution vanishes (terms with antinucleons can be considered similarly). The reason for this 
lies in the fact that we still imply a soft cutoff on the 3-momenta of the 
nucleons within the state $\Omega$ [albeit with the position-dependent ``local 
Fermi momentum'' $k_F$]. On the other hand, the argument of the 
$\delta$-functions in Eq.~(\ref{momentum_rep}) does not vanish in the soft 
momentum region. For example, the solution for $|{\bf l}_1|$ is
\eq\label{l1}
|{\bf l}_1|=\frac{-\Delta|{\bf q^\prime}|\cos\theta\pm
\sqrt{\Delta^2|{\bf q^\prime}|^2\cos^2\theta-(M_p^2 {q^\prime_0}^2-\Delta^2)
({q^\prime_0}^2-|{\bf q^\prime}|^2\cos^2\theta)}}{{q^\prime_0}^2-|{\bf q^\prime}|^2\cos^2\theta}\, ,
\en
where $\theta$ is the angle between ${\bf q^\prime}$ and ${\bf l}_1$, and
\eq\label{Delta}
\Delta=\frac{{q^\prime_0}^2-|{\bf q^\prime}|^2+M_p^2-M_n^2}{2}\, .
\en
The quantity $\Delta$ is of  order $p^2$ in chiral counting.
It is easy to see that, whenever the solution of Eq.~(\ref{l1}) exists,
the quantity $|{\bf l}_1|$ counts at chiral order $1$.
For this reason, the quantity $\Pi^{(2)}(E;{\bf q^\prime},{\bf q})$
vanishes at  $O(p^5)$ in ChPT.

\setcounter{equation}{0}
\section{Equation for the pion-nucleus bound states}
\label{sec:bound}

From equations~(\ref{DbarD}) and (\ref{Pi}) one can define the pion
scattering amplitude $T(E,{\bf q^\prime},{\bf q})$
in the presence of the nucleus
\eq\label{scatT}
D(E,{\bf q^\prime},{\bf q})=\frac{(2\pi)^3\delta^3({\bf q^\prime}-{\bf
q})}{E^2-{\bf q}^2-m_{\pi}^2} 
+\frac{1}{E^2-{\bf q^\prime}^2-m_{\pi}^2}\, T(E,{\bf q^\prime},{\bf q})\,
\frac{1}{E^2-{\bf q}^2-m_{\pi}^2} ,
\en
which obeys  the Lippmann-Schwinger equation
\eq\label{LS}
T(E;{\bf q^\prime},{\bf q})=\Pi(E,{\bf q^\prime},{\bf
q})+\int\frac{d^3{\bf k}}{(2\pi)^3}\, 
\Pi(E,{\bf q^\prime},{\bf k})\frac{1}{E^2-{\bf k}^2-m_{\pi}^2}\, T(E,{\bf k},{\bf q})\, ,
\en
with the self-energy $\Pi(E;{\bf q^\prime},{\bf q})$ playing the role of the potential.
Up to and including $O(p^5)$, this quantity is given by 
\eq\label{Pi5}
&&\Pi(E;{\bf q^\prime},{\bf q})=(2\pi)^3\delta^3({\bf q^\prime}-{\bf
q})\tilde\Pi^{(0)}(E^2-{\bf q}^2)+ 
\Pi^{(1)}(E;{\bf q^\prime},{\bf q})\, ,
\nonumber\\[2mm]
&&\Pi^{(1)}(E;{\bf q^\prime},{\bf q})
=\Pi^{(1\gamma)}+\Pi^{(an)}+\Pi^{(WT)}+\Pi^{(cf)}+\Pi^{(P)}\, ,
\en
where the individual terms of Eq.~(\ref{Pi5}) are given in Eqs.~(\ref{Mplus}), (\ref{P1gamma_r}), (\ref{Pi_an}), (\ref{contact}) and (\ref{Pi-P}), respectively.

The quantity $\tilde\Pi^{(0)}(E^2-{\bf q}^2)$ defined by
Eq.~(\ref{Mplus}) is both infrared-divergent and
gauge-dependent. Below we shall demonstrate that, despite this fact,
the position of the bound-state pole in the scattering amplitude
$T(E;{\bf q^\prime},{\bf q})$ is -- at this order in chiral expansion -- both
infrared-finite and gauge-independent. In order to prove this, we note
that, using the theory of scattering on two potentials, the scattering
amplitude can be given as   
\eq\label{kappanu}
T(E;{\bf q^\prime},{\bf q})&=&(2\pi)^3\delta^3({\bf q^\prime}-{\bf
q})\nu(E^2-{\bf q}^2) \nonumber \\
&&+\kappa^{1/2}(E^2-{\bf q^\prime}^2)\tau(E;{\bf q^\prime},{\bf q})\kappa^{1/2}(E^2-{\bf q}^2) ,
\nonumber\\[2mm]
&&\hspace*{-3cm}
\kappa(s)=[1-A^{(0)}-(s-m_{\pi}^2)B^{(0)}(s)]^{-1}\, ,\quad
\nu(s)=(m_{\pi}^2-s)[1-\kappa(s)]\, ,
\en
where $A^{(0)},~B^{(0)}$ are defined in Eq.~(\ref{Mplus}), and
$\tau(E;{\bf q^\prime},{\bf q})$ 
obeys to the following Lippmann-Schwinger equation
\eq\label{tau}
\tau(E;{\bf q^\prime},{\bf q})&=&U(E,{\bf q^\prime},{\bf
q})+\int\frac{d^3{\bf k}}{(2\pi)^3}\, 
 U(E,{\bf q^\prime},{\bf k})\frac{1}{E^2-{\bf k}^2-m_{\pi}^2}\,
\tau(E,{\bf k},{\bf q})\, , 
\nonumber\\[2mm]
U(E,{\bf q^\prime},{\bf k})&=&\kappa^{1/2}(E^2-{\bf q^\prime}^2)\,
\Pi^{(1)}(E;{\bf q^\prime},{\bf q})\, 
\kappa^{1/2}(E^2-{\bf q}^2)\, . 
\en
From equation (\ref{kappanu}) it is clear that the position of the
poles in $T(E;{\bf q^\prime},{\bf q})$ 
and $\tau(E;{\bf q^\prime},{\bf q})$ coincide. On the other hand, the
difference between the quantities $\Pi^{(1)}(E;{\bf q^\prime},{\bf q})$ and
$U(E;{\bf q^\prime},{\bf q})$ which is due to the factors $\kappa^{1/2}$,
starts at $O(p^6)$ and can be neglected. This is the statement which
we aimed to prove: the pole position is determined from the equation 
(\ref{tau}) whose potential $U(E;{\bf q^\prime},{\bf q})$ is infrared-finite
and gauge-independent 
up to and including $O(p^5)$ in ChPT. Note that, since the potential is Hermitian at this order, the position of the bound-state pole can also be found by solving the equivalent Klein-Gordon equation
\eq\label{Schrodinger}
[E^2-{\bf q^\prime}^2-m_{\pi}^2]\Phi_E({\bf
q^\prime})=\int\frac{d^3{\bf q}}{(2\pi)^3}\, 
U(E;{\bf q^\prime},{\bf q})\, \Phi_E({\bf q})\, ,
\en
where $\Phi_E({\bf q})$ stands for the wave function of the bound
state in the momentum representation. 

We finally collect all terms of the potential at this
order (replacing $\sigma^{[0,3]}({\bf x})$ 
by $\rho^{[0,3]}({\bf x})$, see Eq.~(\ref{rho-sigma})):
\eq\label{U}
&&U(E;{\bf q^\prime},{\bf q})=\int d^3{\bf x}\,{\rm e}^{-i({\bf q^\prime}-{\bf
q}) \cdot {\bf x}}\, 
[\tilde U(E;{\bf q^\prime},{\bf q};{\bf x})+O(p^6)]\, ,
\\[2mm]
&&\tilde U(E;{\bf q^\prime},{\bf q};{\bf x})=-\int d^3{\bf
r}\,\frac{\alpha E}{|{\bf x}-{\bf r}|}\, 
[\rho^0({\bf r})+\rho^3({\bf r})]
\nonumber\\[2mm]
&-&
\frac{E-e^2F^2f_2}{2F^2}\,\rho^3({\bf x})
+\frac{4c_1m_{\pi^0}^2-2(c_2+c_3)E^2+2c_3{\bf q^\prime} \cdot {\bf q}+2e^2F^2f_1}{F^2}\,\rho^0({\bf x})
\nonumber\\[2mm]
&+&\frac{g_A^2}{2F^2}\, \biggl\{
\biggl[\frac{E^2-2{\bf q^\prime} \cdot {\bf q}}{2M_N}
+\frac{3({\bf q^\prime} \cdot {\bf q})^2-{\bf q^\prime}^2{\bf
q}^2}{4M_NE^2}\biggr]\rho^0({\bf x}) 
\nonumber \\[2mm]
&&+\frac{{\bf q^\prime}\cdot  
{\bf q}}{E}\,\biggl(1+\frac{M_n-M_p}{E}\biggr)
\rho^3 ({\bf x})\biggr\}\, .
 \label{Utwiggle}
\en
Equation~(\ref{Utwiggle}) represents our main result: the
complete expression of the optical potential for pion scattering on
the finite nucleus up to and including $O(p^5)$ in ChPT in the
presence of electromagnetic interactions and strong isospin-breaking
effects. Note that, unlike most of previous descriptions,
 the framework described in the present paper enables
one to unambiguously obtain the explicit dependence of the optical
potential  on the momenta (off-shell) 
${\bf q}$ and ${\bf q^\prime}$. For example, the
terms proportional to  
$({\bf q^\prime} \cdot {\bf q})^2-{\bf q^\prime}^2{\bf q}^2$ vanish in
the infinite 
medium. They cannot be obtained from an extrapolation of standard
in-medium ChPT or semi-phenomenological approaches starting from pion
scattering in nuclear matter.

\setcounter{equation}{0}
\section{Comparison with  existing approaches}
\label{sec:comp}

In this section, we compare our results 
\begin{itemize}
\item[i)-iii)]
to the existing calculations of the in-medium pion mass shift in ChPT~\cite{KW,Annals,Korea};
\item[iv)]
to the pion-nucleus optical potential obtained in Ref.~\cite{Kolo} from ChPT;
\item[v)]
to the empirical pion-nucleus optical potentials which were widely used in the literature to describe deeply bound states of pions on the heavy nuclei~\cite{potential}.
\end{itemize}

\noindent
i) In order to compare with the calculations performed in the infinite nuclear 
medium,
we put $\rho^0({\bf x})=\rho_p+\rho_n$, $\rho^3({\bf x})=\rho_p-\rho_n$, where
$\rho_{p,n}=\frac{1}{3\pi^2}\,(k_F^{(p,n)})^3$ and $k_F^{(p,n)}$ stands for 
the Fermi-momentum of the proton (neutron). 
In the limit of a uniform medium, our expression for the self-energy 
 in the absence of electromagnetic interactions ($e=0$) coincides with
the expression for the same quantity at $O(p^5)$,  
which is given
in Eq.~(4.5) of Ref.~\cite{Annals}, except for the term proportional to
$g_A^2(M_n-M_p)$ (or, equivalently, $g_A^2c_5B(m_d-m_u)$ 
in the chiral expansion). Note that, since $e=0$ is 
assumed in Ref.~\cite{Annals}, at $O(p^5)$ there 
is no difference whether the LEC $c_1$ is multiplied by $m_{\pi}^2$ (as 
in \cite{Annals}) or $m_{\pi^0}^2$ (as in the present paper). For the same 
reason, the contribution 
involving the  LECs $f_1,~f_2$ is not present in 
Eq.~(4.5) of Ref.~\cite{Annals}.

\bigskip

\noindent ii)
In Ref.~\cite{KW} the threshold self-energy ($E=m_\pi$) is evaluated
at $O(p^6)$ in ChPT, 
at vanishing 3-momentum,
and with $e=0$, $m_u=m_d$, but at $k_F^{(p)}\neq k_F^{(n)}$. The
result, given by Eqs. (4)-(5) of this paper, does not depend on the energy
$E$ as 
well. Up to and including order $p^5$, this result agrees 
(apart from the terms proportional to $e^2$ and $m_d-m_u$)
with our result, if one sets $E^2=m_\pi^2$.
However, ref.~\cite{KW} does include very important double scattering
contributions which first appear at $O(p^6)$.

\bigskip

\noindent iii)
The results obtained in Ref.~\cite{Korea} are similar to those of 
Ref.~\cite{KW}, with the exception that in Ref.~\cite{Korea} the restriction
$E^2=m_\pi^2$ is removed. As was mentioned above, both \cite{KW,Korea}
determine the pole position from the two-point function of the interpolating
pion fields. In this quantity one encounters the so-called off-shell ambiguity
for $E^2\neq m_\pi^2$,
which should disappear in the calculation of the pole position in order to be 
compatible with general
 principles  of quantum field theory.
In Ref. \cite{Korea} it is demonstrated that this actually happens if one
expands the pole position in powers of the quark mass and eliminates
the contributions at $O(p^7)$ and higher (see also \cite{Kubodera}). 

\bigskip

Concerning numerical studies of the in-medium pion mass shift, the   
perturbative estimates in all three papers \cite{KW,Annals,Korea}
yield results around  
$10~{\rm MeV}$, which is too small as compared to the ``empirically'' 
deduced  shift
$\sim 25~{\rm MeV}$. In Ref. \cite{Annals} it is shown, that making the partial
resummation of the higher-order terms, 
it is possible to obtain the result for the mass shift which differs from the
``perturbative'' solution by almost a factor of 2. The main reason for this 
difference is the  strong energy-dependence of the self-energy
operator.
It should be noted, of course, that the concept of a pion mass shift
in nuclear matter, while of some theoretical interest, is of only
limited relevance in the context of Coulomb-bound pionic atom
states. It is in fact more satisfactory to directly test
the theoretical predictions for the bound-state observables on the experiment
by solving the wave equation self-consistently for the
finite system -- e.g., as done in Ref.~\cite{Kolo}.

\bigskip

\noindent
iv) In order to compare with the pion-nucleus optical potential
obtained in Ref.~\cite{Kolo} by extrapolating the results of
in-medium ChPT to the finite nuclei, we rewrite Eq.~(\ref{U})   
as follows:
\eq\label{Cspd}
\tilde U(E;{\bf q^\prime},{\bf q};{\bf x})&=&U_C(E;{\bf x})+U_S(E;{\bf x})
-({\bf q^\prime} \cdot {\bf q})U_P(E;{\bf x})
\nonumber\\[2mm]
&+&(3({\bf q^\prime} \cdot {\bf q})^2-{\bf q^\prime}^2{\bf q}^2)U_D(E;{\bf x})\, ,
\en
where $U_C(E;{\bf x})$ is the Coulomb term (first line of
Eq.~(\ref{Utwiggle})), and $S-,P-,D-$ wave parts are given by 
\eq\label{Uspd}
U_S(E;{\bf x})\!&=&\!
-\frac{E-e^2F^2f_2}{2F^2}\,\rho^3({\bf x})
\nonumber\\[2mm]
&+&\frac{4c_1m_{\pi^0}^2-2(c_2+c_3-g_A^2/8M_N)E^2+2e^2F^2f_1}{F^2}\,\rho^0({\bf x})
\nonumber\\[2mm]
U_P(E;{\bf x})\!&=&\!-\frac{2(c_3-g_A^2/4M_N)}{F^2}\,
\rho^0({\bf x}) 
+\frac{g_A^2}{2F^2E}\,\biggl[1+\frac{M_n-M_p}{E}\biggr]\rho^3({\bf x})\, ,
\nonumber\\[2mm]
U_D(E;{\bf x})\!&=&\!\frac{g_A^2}{8F^2M_NE^2}\,\rho^0({\bf x})\, .
\en
\begin{itemize}

\item
The $S$-wave part of the optical potential coincides with the one of
Ref.~\cite{Kolo} at $O(p^5)$ in ChPT when setting $e=0$. The
low-energy constant $c_1$ is related to the pion-nucleon sigma term,
$\sigma_N=-4 c_1 m^2_{\pi}$. The combination $c_2+c_3$ receives an
important contribution from the $\Delta$ resonance in the $P$-wave
$\pi N$ amplitude. The values of these LEC's are individually much
larger that their combination appearing in $U_S(m_\pi;{\bf x})$: they
must conspire in just such a way as to reproduce the observed
approximate vanishing of the isospin-even $\pi N$ amplitude at
threshold. A small departure from this subtle balance at threshold is
expected to have a large impact on pionic bound state energies. This
is pointed out in Ref.~\cite{Kolo} where it is argued that a large
part of the ``missing $S$-wave repulsion'' observed in pionic atoms
can be explained by the strong energy dependence of the low-energy
$\pi N$ amplitude in ChPT, while the proper gauge-invariant inclusion
of the Coulomb potential accounts for most of the remaining
effect. This calculation also includes the important double scattering 
terms of $O(p^6)$.
The result, concerning the energy-dependence at $O(p^5)$, 
is consistent with the findings of
Ref.~\cite{Annals} for the infinite medium.

Finally we note that the electromagnetic corrections represented by
the LECs $f_1$ and $f_2$ were so far not taken into account in the
literature.

\item
For the $P$-wave part of the optical potential, the authors of 
Ref.~\cite{Kolo} use the time-honored phenomenological
parameterization which systematically reproduces a large amount of
pionic atom data. This parameterization includes effects of higher
order in density, such as the Lorentz-Lorentz correction and pion
absorption \cite{Ericson,EW}, which cannot be handled within the
one-nucleon sector of ChPT. On the other hand, \NChPT makes rigorous
leading-order statements concerning the $P$-wave pion self energy for
finite systems, and it is at this level where a comparison with
phenomenology is useful.
In particular $U_P(E;{\bf 
x})$ can be compared to the leading (linear in density) term of the 
phenomenological parameterization, represented by the $\pi N$ scattering
volumes    
$\tilde c_0$ and $\tilde c_1$ \cite{Ericson,EW}. (The scattering
volumes are usually denoted by $c_0$ and $c_1$, but we need to avoid
confusion with the LECs $c_1,...,c_3$.)  
One may replace $E\to m_{\pi}$ 
in our expression, and rewrite the $P$-wave part of the optical potential as
\eq\label{UP}
\frac{1}{4\pi}\,\biggl(1+\frac{m_{\pi}}{M_N}\biggr)U_P(m_{\pi};{\bf x})
=\tilde c_0\rho^0({\bf x})-\tilde c_1\rho^3({\bf x})\, .
\en 
Using the empirical value $\tilde c_0 =0.21 m_{\pi}^{-3}$, from 
Table~6.2 of 
Ref. \cite{EW}, $g_A=1.267$ and
$F=92.4$~MeV, we obtain $c_3=-3.2$~GeV$^{-1}$.
There is  a theoretical uncertainty in
this number, given by the typical size of next-to-leading order
corrections. As a consequence, our leading-order
estimate of the LEC $c_3$ in the low-energy $\pi N$ effective
Lagrangian is not incompatible with the value $c_3=-4.7~{\rm
GeV}^{-1}$  quoted in Ref. \cite{Annals}.  As
for the isovector term, we find $\tilde 
c_1=0.17~m_{\pi}^{-3}$  at $E=m_{\pi}$, in agreement with the value
used in the set~A  of Ref. \cite{EW}.  

Note that the relation between the LEC $c_3$ and the phenomenological
$P$-wave scattering volume $\tilde c_0$ is different in the case of
scattering on a single nucleon. Indeed, it is well known that the
contribution from the Born (nucleon pole) diagrams to this quantity is
zero up to and 
including terms of $O(1/M_N)$ \cite{BKM97}. The differences between the
two cases originate in recoil effects which differ for a single
nucleon and for the nucleus as a whole.
In the case of scattering on a nucleon the
recoil is an effect of order $O(p)$. In the case of a nucleus it is
suppressed by an additional factor of $1/A$. It is therefore natural to
find such  differences already in our $O(p^5)$ calculation. These terms are
corrections to the low-density theorem relating the pion self-energy to
the forward $\pi N$ scattering amplitude. It is not possible to identify
such corrections in the conventional in-medium ChPT, since the
restriction to 
${\bf p}={\bf q}$ makes the  separation between $P$-wave and $S$-wave
ambiguous.

\item
There is no $D$-wave part in the optical potential of Ref.~\cite{Kolo}. 
In our approach, 
this contribution is partly due to the finite-size effect.
As  already mentioned, the presence of such 
terms cannot be reliably established from
the extrapolation of the result obtained at uniform density. They can
only be controlled in 
the consistent approach described in this paper.
\end{itemize}

\bigskip

\noindent
v) The phenomenological optical potential for the pion-nucleus
interaction close to threshold  is commonly written in coordinate space  as
\cite{potential} 
\eq\label{opt}
2m_{\pi}\tilde U^{\rm phen}=q(r)+\nabla \alpha(r) \nabla\, ,
\en
where $q(r)$ and $\alpha(r)$ are taken to be energy-independent. From
Eq.~(\ref{Uspd}) we observe that there are more terms in the result
obtained within ChPT: 
for example, we have in addition the $D$-wave part of the potential
which is absent in Eq.~(\ref{opt}). What is important, the
counterparts of $q(r),~\alpha(r)$ in ChPT depend also explicitly on the energy
$E$. This energy dependence stems from the underlying chiral
pion-nucleon  dynamics, and is uniquely determined in ChPT.
 
\bigskip

A few final remarks are in order. First, we wish to mention that in 
Refs.~\cite{potential,KW,Annals,Korea,Kolo} 
the terms with $e\neq 0$, $m_d-m_u\neq 0$ in the non-Coulomb part of the
optical potential have always been  neglected.
In this paper, we have included them on the same footing as all other
(strong) terms that arise at the same chiral order.
Note that the additional electromagnetic terms have sometimes led to
significant corrections in the observables bound states of hadrons
(see e.g.~\cite{EPJC,atom}). 

As already mentioned, the optical potentials given in
Refs.~\cite{potential,Kolo} 
include additional terms, non-linear in the baryon density. 
It is well known that these terms give important 
contributions to the binding energies and widths of pionic atoms. In our
counting all these terms start at $O(p^ 6)$ and have not (yet) been
considered  in the present paper. Clearly,
 in order to achieve a good description of pionic atoms, one
has to carry out the calculations, outlined in this paper, at least up
to and including $O(p^6)$.

The focus in this paper is on the presentation of the \NChPT framework
for finite systems. Numerical algorithms for solving the bound-state
equation and finding the systematic ChPT expansion of the pole
positions are a different issue which deserves a separate study. Such
an investigation is highly non-trivial. It requires checking the
convergence of the chiral expansion for the eigenvalues and examining
the stability of these eigenvalues with respect to small variations of
the self-energy operator, bearing in mind the strong energy dependence
discussed previously.

\setcounter{equation}{0}
\section{Conclusions}
\label{sec:cncl}

\begin{itemize}

\item[i)]
In this paper we propose a novel approach to construct ChPT in the
background of a {\em finite} 
nucleus: ``Chiral 
Perturbation Theory for Heavy Nuclei'' (\NChPT). It develops the rules
for systematically evaluating the pion-nucleus optical potential
directly for a finite size system. In the present paper the full
result of the calculations at $O(p^5)$ is presented. 
\item[ii)]
The approach  is based on certain 
approximations concerning the description of the nucleus containing $A$ 
nucleons, where $A$ is a large number.
The content of these approximations, given by Eq.~(\ref{ME_Shar}), is
as follows: 
 the nucleus in 
\NChPT does not contain free pions and photons. Secondly,
 the nucleus is not excited by the external pion bound in the atom.
\item[iii)]
For simplicity, it is  also assumed that the nucleus is static
and in a spin-saturated ground state. These assumptions are purely
technical and -- if needed -- can be released at a later stage. 
\item[iv)]
In \NChPT,
the scattering amplitude of the pion in the nuclear background
is given by a sum of
 terms, each of which is a product of two factors: the scattering
amplitude of the pion on $0,1,2,\cdots$ 
nucleons which is systematically evaluated in ChPT, and the matrix
elements of the normal products of the free fermionic fields between
the nuclear states. These (and only these) matrix elements summarize
the necessary  nuclear structure information. 
\item[v)]
A crucial ingredient of our approach is the
chiral counting for the nuclear matrix elements. The bookkeeping based on
chiral symmetry allows one to carry out a Taylor expansion of these matrix
elements. This expansion drastically reduces the sensitivity of the final
results on the nuclear input.  
\item[vi)]
The approach can be pursued systematically to higher chiral orders:

First, the chiral counting allows  
one to unambiguously organize different contributions, and 
off-shell ambiguities are absent from the beginning. 

There are no ultraviolet 
divergences arising at any step, which would signal an internal
inconsistency: the coefficient functions $S^{(n)}$ are finite by
construction, and the phenomenological input on the nuclear matrix
elements  does not introduce ultraviolet divergences either. 

The only general property of the theory which might be lost in the 
approximations that have led to \NChPT, is the gauge invariance of the
bound-state energy at high chiral orders. We have checked gauge invariance
explicitly at $O(p^5)$. However, at the present stage, 
we are not aware of a general 
proof that this can be done in all orders. Explicit gauge dependence 
 appearing at $O(p^x)$ would signal that, starting from this order,
one cannot neglect components of the nuclear wave function that
contain pions and photons. 
From the phenomenological point of view, the situation
is harmless.

\item[vii)]
Our approach is the generalization of in-medium ChPT which is
based on the Fermi-gas model. In the limit of uniform density, our
approach reduces to the conventional in-medium ChPT. 

\item[viii)]
In this paper  we present a systematic
derivation of the leading-order  pion-nucleus optical potential in ChPT,
including isospin-breaking contributions, and provide the detailed
comparison to other results existing in the literature. It is
demonstrated that the present approach generates more terms, with
higher spatial derivatives, than conventional models using the local
density approximation. 
In particular, this concerns an additional contribution in the $P$-wave
scattering volume, as well as a $D$-wave term which is absent
in the conventional parameterizations and can be interpreted as a genuine
 finite-size effect.
The negligence of isospin-breaking contributions cannot be
justified since these come at the same chiral order as the strong
contributions, and since one anyway includes some of the
electromagnetic contributions -- e.g. the Coulomb potential. An
accurate fit to the measured bound-state energies of  
pionic atoms might help to set bounds on the value of the electromagnetic
low-energy constant $f_1$ which enters the expression of the optical
potential at $O(p^5)$.  
\item[ix)]
Carrying out complete calculations at $O(p^6)$ is challenging from
several points of view. First of all, the contributions at this order,
including the contributions which are quadratic in the baryon density,
are known to be  important. We do not
give any numerical estimates in this paper, based on the $O(p^5)$
calculations alone. 
We plan to address $O(p^6)$ calculations, as well as the
thorough numerical analysis of the bound-state problem,  in  future
publications.  

\end{itemize}

{\em Acknowledgments:}
We thank Torleif Ericson, J\"{u}rg Gasser, George Jackeli, 
Norbert Kaiser, Evgeni Kolomeitsev, Ulf Mei{\ss}ner, Jose Oller, 
Georges Ripka and Andreas Wirzba for discussions.

\appendix

\renewcommand{\thesection}{\Alph{section}}
\renewcommand{\theequation}{\Alph{section}\arabic{equation}}

\setcounter{equation}{0}
\section{Chiral Lagrangians}
\label{app:lagr}

For convenience, in this appendix we collect chiral Lagrangians, which are
used in the calculation of the pion self-energy at $O(p^5)$. Notations and
conventions are identical to the used in Ref.~\cite{EPJC}, from which the 
formulae below are taken
\eq\label{lagr-init}
&&{\cal L}_\gamma+{\cal L}_\pi^{(p^2)}+{\cal L}_\pi^{(e^2)}=
-\frac{1}{4}\,F_{\mu\nu}F^{\mu\nu}-\frac{1}{2\xi}\,(\partial_\mu A^\mu)^2
\nonumber\\[2mm]
&&+\frac{F^2}{4}\,\langle d^\mu U^\dagger d_\mu U+\chi^\dagger U
+U^\dagger\chi\rangle +ZF^4\langle {\cal Q} U{\cal Q}U^\dagger\rangle\, ,
\nonumber\\[2mm]
&&{\cal L}_\pi^{(p^4)}=\sum_{i=1}^{7}l_iO_i^{(p^4)}\, ,\quad\quad
{\cal L}_\pi^{(e^2p^2)}=F^2\sum_{i=1}^{10}k_iO_i^{(e^2p^2)}\, ,
\nonumber\\[2mm]
&&{\cal L}_N^{(p)}=\bar\psi (i\not\! D-m+\frac{1}{2}\, g_A\not\! u\gamma_5)\psi\, ,
\nonumber\\[2mm]
&&{\cal L}_N^{(p^2)}=\sum_{i=1}^{7}c_i\bar\psi O_i^{(p^2)}\psi\, ,\quad\quad
{\cal L}_N^{(e^2)}=F^2\sum_{i=1}^{3}f_i\bar\psi O_i^{(e^2)}\psi\, .
\en
Here, $\xi$ denotes the gauge parameter ($\xi=1$ in the Feynman gauge), and
$\langle(\cdots)\rangle$ stands for the trace over the isospin indices.
The building blocks for constructing the Lagrangian are
\eq\label{building}
&&d_\mu U=\partial_\mu U-i{\cal R}_\mu U+iU{\cal L}_\mu\, ,\quad
\pmatrix{{\cal R}_\mu\cr\cal{L}_\mu}=v_\mu\pm a_\mu+{\cal Q}A_\mu\, ,
\nonumber\\[2mm]
&&{\cal R}_{\mu\nu}=\partial_\mu {\cal R}_\nu-\partial_\nu 
{\cal R}_\mu-i[{\cal R}_\mu,{\cal R}_\nu]\, ,\quad
{\cal L}_{\mu\nu}=\partial_\mu {\cal L}_\nu-\partial_\nu 
{\cal L}_\mu-i[{\cal L}_\mu,{\cal L}_\nu]\, ,
\nonumber\\[2mm]
&&\chi = 2 B (s+ip)\, ,\quad
d_\mu \chi=\partial_\mu \chi-i{\cal R}_\mu \chi+i\chi{\cal L}_\mu\, ,\quad
F_{\mu\nu}=\partial_\mu A_\nu-\partial_\nu A_\mu\, ,
\nonumber\\[2mm]
&&{\cal C}_R^\mu=-i[{\cal R}_\mu,{\cal Q}]\, ,\quad
{\cal C}_L^\mu=-i[{\cal L}_\mu,{\cal Q}]\, ,\quad
{\cal Q}=e\,{\rm diag}(\frac{2}{3},-\frac{1}{3})\, ,
\en
in the meson sector, and 
\eq\label{blocks}
&&D_\mu=\partial_\mu+\Gamma_\mu\, ,\quad
U=u^2\, ,
\quad u_\mu=iu^\dagger d_\mu Uu^\dagger\, ,
\nonumber\\[2mm]
&&\Gamma_\mu=\frac{1}{2}\,[u^\dagger,\partial_\mu u]
-\frac{i}{2}\, u^\dagger R_\mu u
-\frac{i}{2}\, u L_\mu u^\dagger\, ,
\nonumber\\[2mm]
&&\chi_\pm=u^\dagger\chi u^\dagger\pm u\chi^\dagger u\, ,\quad
\hat\chi_+=\chi_+-\frac{1}{2}\,\langle\chi_+\rangle\, ,
\nonumber\\[2mm]
&&F^\pm_{\mu\nu}=u^\dagger R_{\mu\nu}u\pm uL_{\mu\nu}u^\dagger\, ,\quad
\hat F^+_{\mu\nu}=F^+_{\mu\nu}-\frac{1}{2}\,\langle F^+_{\mu\nu}\rangle\, ,
\nonumber\\[2mm]
&&Q_\pm=\frac{1}{2}\,(uQu^\dagger\pm u^\dagger Qu)\, ,\quad
\hat Q_\pm=Q_\pm-\frac{1}{2}\,\langle Q_\pm\rangle\, ,\quad
Q=e\,{\rm diag}(1,0),
\en
in the nucleon sector. Further,
$R_\mu$, $L_\mu$, $R_{\mu \nu}$, $L_{\mu \nu}$ are defined just like their 
pionic counterparts ${\cal R}_\mu$, ${\cal L}_\mu$, ${\cal R}_{\mu \nu}$, 
${\cal L}_{\mu \nu}$ respectively, with ${\cal Q}$ replaced by $Q$.
As usual,  $j\doteq\{s,p,v_\mu,a_\mu\}$ denote external scalar, pseudoscalar, vector and axial fields, in the form of $2\times 2$ matrices.
In all expressions we drop the terms which do not contain pion fields, and terms of order $e^4$.

\begin{table}
\caption{
Operator basis and the divergent parts (in the Feynman gauge) of the LECs in 
the $O(p^4)$ meson Lagrangian \cite{GL}
and $O(e^2p^2)$ meson Lagrangian \cite{KU}. 
The terms that do not contain pion fields, 
and terms of order $e^4$ are not displayed. }
\label{tab:Lmeson}

\begin{center}
\def\arraystretch{1.4}
\begin{tabular}{|r|c|c|}
\hline\hline
$i$ & $O_i^{(p^4)}$ & $\gamma_i$ \\
\hline
$1$&$\frac{1}{4}\,\langle d^\mu U^\dagger d_\mu U\rangle^2$ & $\frac{1}{3}$\\
$2$&$\frac{1}{4}\,\langle d^\mu U^\dagger d^\nu U\rangle\langle d_\mu
U^\dagger d_\nu U\rangle$ & $\frac{2}{3}$\\
$3$&$\frac{1}{16}\,\langle\chi^\dagger U+U^\dagger\chi\rangle^2$ & $-\frac{1}{2}$\\
$4$&$\frac{1}{4}\,\langle d^\mu U^\dagger d_\mu\chi+d^\mu\chi^\dagger d_\mu
U\rangle$ & $2$\\ 
$5$&$\langle {\cal R}^{\mu\nu}U{\cal L}_{\mu\nu}U^\dagger\rangle$ & 
$-\frac{1}{6}$\\
$6$&$\frac{i}{2}\,\langle {\cal R}^{\mu\nu}d_\mu Ud_\nu U^\dagger$ 
& $-\frac{1}{3}$\\
& $+{\cal L}^{\mu\nu}d_\mu U^\dagger d_\nu U\rangle$ & \\
$7$&$-\frac{1}{16}\,\langle\chi^\dagger U-U^\dagger\chi\rangle^2$ & $0$\\
\hline\hline
$i$ & $O_i^{(e^2p^2)}$ & $\sigma_i$ \\
\hline
$1$&$\langle d^\mu U^\dagger d_\mu U\rangle\langle{\cal Q}^2\rangle$ & $-\frac{27}{20}-\frac{1}{5}\,Z$\\
$2$&$\langle d^\mu U^\dagger d_\mu U\rangle
\langle{\cal Q}U{\cal Q}U^\dagger\rangle$ & $2Z$\\
$3$&$\langle d^\mu U^\dagger{\cal Q}U\rangle
\langle d_\mu U^\dagger{\cal Q}U\rangle$ & $-\frac{3}{4}$
\\
&$+\langle d^\mu U{\cal Q}U^\dagger\rangle
\langle d_\mu U{\cal Q}U^\dagger\rangle$ & \\
$4$&$\langle d^\mu U^\dagger{\cal Q}U\rangle
\langle d_\mu U{\cal Q}U^\dagger\rangle$ & $2Z$\\
$5$&$\langle\chi^\dagger U+U^\dagger\chi\rangle\langle{\cal Q}^2\rangle$ & $-\frac{1}{4}-\frac{1}{5}\,Z$\\
$6$&$\langle\chi^\dagger U+U^\dagger\chi\rangle
\langle{\cal Q}U{\cal Q}U^\dagger\rangle$ & $\frac{1}{4}+2Z$ \\
$7$&$\langle(\chi U^\dagger+U\chi^\dagger){\cal Q}$
& $0$ \\
&$+(\chi^\dagger U+U^\dagger\chi){\cal Q}\rangle\langle{\cal Q}\rangle$ & \\
$8$&$\langle(\chi U^\dagger-U\chi^\dagger){\cal Q}U{\cal Q}U^\dagger$
&$\frac{1}{8}-Z$ \\
&$+(\chi^\dagger U-U^\dagger\chi){\cal Q}U^\dagger{\cal Q}U\rangle$ & 
\\
$9$&$\langle d_\mu U^\dagger[{\cal C}_R^\mu,{\cal Q}]U+
d_\mu U[{\cal C}_L^\mu,{\cal Q}]U^\dagger\rangle$ & $\frac{1}{4}$\\
$10$&$\langle {\cal C}_R^\mu U{\cal C}_{L\mu}U^\dagger\rangle$ & $0$\\
\hline
\end{tabular}
\end{center}
\end{table}

\begin{table}
\caption{
Operator basis in the $O(p^2)$ pion-nucleon Lagrangian
\cite{Fettes} and in the $O(e^2)$ pion-nucleon Lagrangian \cite{Muller}.}
\label{tab:Lnucleon}

\begin{center}
\def\arraystretch{1.4}
\begin{tabular}{|r|c|c|}
\hline\hline
$i$ & $O_{i}^{(p^2)}$ & $O_{i}^{(e^2)}$\\
\hline
$1$&$\langle\chi_+\rangle$ & $\langle\hat Q_+^2-Q_-^2\rangle$ \\
$2$&$-\frac{1}{4m^2}\,\langle u_\mu u_\nu\rangle
(D^\mu D^\nu+\mbox{h.c.})$ & $\langle Q_+\rangle\hat Q_+$ \\
$3$&$\frac{1}{2}\,\langle u_\mu u^\mu\rangle$ 
& $\langle\hat Q_+^2+Q_-^2\rangle$ \\
$4$&$\frac{i}{4}\,\sigma^{\mu\nu}[u_\mu,u_\nu]$&\\
$5$&$\hat\chi_+$&\\
$6$&$\frac{1}{8m}\,\sigma^{\mu\nu}F^+_{\mu\nu}$&\\
$7$&$\frac{1}{8m}\,\sigma^{\mu\nu}\langle F^+_{\mu\nu}\rangle$&\\
\hline 

\end{tabular}
\end{center}

\end{table}

In the above formulae, the symbol $e$ stands for the electric charge. 
The quantities $F$, $m$ and $g_A$ are the pion decay constant, nucleon mass
and the nucleon axial constant in the chiral limit. The quantity $B$ is 
related to the quark condensate in a standard manner, and the quantity
$Z$ is expressed through the charged-neutral pion mass difference in
the chiral limit
\eq\label{diff}
m_{\pi}^2-m_{\pi^0}^2=2e^2F^2Z+\cdots
\en
Further, the LECs $l_i$ and $k_i$ are ultraviolet-divergent
\eq\label{lk_div}
l_i=\gamma_i\lambda+l_i^r(\mu)\, ,\quad
k_i=\sigma_i\lambda+k_i^r(\mu)\, ,
\en
where
\eq\label{lambda}
\lambda=\frac{\mu^{d-4}}{16\pi^2}\,\biggl(\frac{1}{d-4}-\frac{1}{2}\,
[\Gamma'(1)+\ln 4\pi+1]\biggr)\, .
\en

The components of the external sources are defined as
\eq\label{external}
v_\mu=\frac{1}{2}\,v_\mu^0+\frac{\tau^n}{2}\,v_\mu^n\, ,\quad
a_\mu=\frac{\tau^n}{2}\,a_\mu^n\, ,\quad
s=s^0+\tau^ns^n\, , \quad
p=\tau^np^n\, .
\en

In tables~\ref{tab:Lmeson} and~\ref{tab:Lnucleon}, we collect the operator
basis for meson and meson-nucleon Lagrangians. In addition, in 
table~\ref{tab:Lmeson}, the divergent parts of the LECs
$l_i$, $k_i$ are listed (see Eq.~(\ref{lk_div})).

\end{document}